\documentclass[a4paper,fleqn,usenatbib]{mnras}
\usepackage{newtxtext,newtxmath}

\usepackage[T1]{fontenc}
\usepackage{ae,aecompl}

\usepackage{graphicx}	
\usepackage{amsmath}	
\usepackage{amssymb}	
\usepackage{lscape}
\usepackage{color}

\newcommand{\kep}{\textit{Kepler} }

\newcommand{\todcor}{\textsc{todcor} }

\newcommand{\jkt}{\textsc{jktebop} }
\newcommand{\jktabs}{\textsc{jktabsdim} }

\newcommand{\ms}{m~s$^{-1}$}

\title[HIDES spectroscopy of \kep DEBs]{HIDES spectroscopy of bright detached eclipsing
binaries from the \kep field -- II. Double- and triple-lined objects.
}
\author[K. G. He{\l}miniak et al.]
{K. G. He{\l}miniak$^{1,4}$\thanks{Subaru Research Fellow, e-mail: xysiek@naoj.org},
N. Ukita$^{2,3}$,
E. Kambe$^{2}$,
S. K. Koz{\l}owski$^{4}$,
P. Sybilski$^{4}$,
\and
H. Maehara$^{2}$,
M. Ratajczak$^{5,4}$,
M. Konacki$^{4}$,
R. K. Paw{\l}aszek$^{4}$
\\
$^{1}$Subaru Telescope, National Astronomical Observatory of Japan, 650 North Aohoku Place, Hilo, HI 96720, USA\\
$^{2}$Okayama Astrophysical Observatory, National Astronomical Observatory of Japan, 3037-5 Honjo, Kamogata, Asakuchi,\\Okayama 719-0232, Japan\\
$^{3}$The Graduate University for Advanced Studies, 2-21-1 Osawa, Mitaka, Tokyo 181-8588, Japan\\
$^{4}$Nicolaus Copernicus Astronomical Center, Department of Astrophysics, ul. Rabia\'nska 8, 87-100 Toru\'n, Poland\\
$^{5}$Instytut Astronomiczny, Uniwersytet Wroc{\l}awski, ul. Kopernika 11, 51-622 Wroc{\l}aw, Poland\\
}

\date{Accepted XXX. Received YYY; in original form ZZZ}

\pubyear{2017}

\begin{document}
\label{firstpage}
\pagerange{\pageref{firstpage}--\pageref{lastpage}}
\maketitle

\begin{abstract}
We present the results of our spectroscopic observations of eight 
detached eclipsing binaries (DEBs), selected from the \kep Eclipsing Binary
Catalog. Radial velocities (RVs) were calculated from 
high resolution spectra obtained with the HIDES spectrograph, attached 
to the 1.88-m telescope of the Okayama Astrophysical Observatory, 
and were used to characterize the targets in combination with the \kep light 
curves. For each binary we obtained a full set of orbital and physical parameters, 
reaching precision below 3\% in masses and radii for 5 pairs. By comparing
our results with theoretical models, we assess the distance, age and evolutionary
status of the researched objects. We also study eclipse timing variations of selected
objects, and identify a new system with a $\gamma$Dor pulsator.
Two systems are triples, and show lines coming from three
components. In one case the motion of the outer star and the perturbation in 
the RVs of the inner binary are clearly visible and periodical, which allows us to
directly calculate the mass of the third star, and inclination of the outer orbit.
In the second case we only see a clear motion of the tertiary, and investigate two
scenarios: that it is a linear trend coming from the orbital motion around the inner
binary, and that it is caused by a planetary mass companion.
When possible, we also compare our results with the literature, and conclude that 
only by combining photometry with RVs it is 
possible to obtain correct physical parameters of both components of a DEB.
\end{abstract}


\section{Introduction}
The field of stellar astrophysics has benefited from the launch of space missions
dedicated for precise photometry, like MOST, CoRoT, or {\it Kepler}. High-precision
light curves (LCs), of a quality not possible to obtain before from the ground, have
revolutionized such branches of astrophysics as asteroseismology or eclipsing binaries.
The latter are one of the most important objects in astronomy, as they allow for direct
determination of parameters like mass and radius, which are very difficult or 
impossible to obtain with a different method. Such a knowledge is the basis for 
further studies of, for example, stellar structure and evolution theory, population
synthesis, galactic archaeology, cosmic distance scale, extrasolar planets, etc. 
\citep{tor10}. No wonder that for decades researchers were interested in obtaining 
very precise and accurate basic stellar parameters. It is now believed that
the results (e.g. masses and radii) are useful for the purposes of modern astrophysics
when they are determined with the precision of 2-3 per cent or better \citep{las02,tor10,sou15}.
To reach it, one needs high quality spectroscopic and photometric data. The former
is usually obtained with stable high-resolution echelle spectrographs, while the
source of the most precise photometry was the original \kep mission. 

This paper is a continuation of the research on bright \kep DEBs, some of 
which have been described in \citet{hel15b} and \citet[][hereafter Paper~I]{hel16}.
The latter work presents the whole observing program in more details. Here
we focus on double- and triple-lined systems. Section~\ref{sec_targets} 
describes the objects, Section~\ref{sec_data} presents the data and methodology,
and results are summarised in Section~\ref{sec_results}, followed by a discussion
in Section~\ref{sec_discus}. 

\section{Targets}\label{sec_targets}

\begin{table*}
\centering
\caption{KEBC information about the observed targets, and their character}\label{tab_cat}
\begin{tabular}{llllllllrc}
\hline
\hline
KIC & KOI & Other name & RA (deg) & DEC (deg) & $P$~(d)$^a$ & $T_0$ (BJD-2450000)$^a$ & $T_{eff}$ & $kmag$ & 3rd?$^b$ \\
\hline
06525196 & 5293 & TYC 3143-604-1 & 292.7180 & 41.9225 &  3.420604  & 4954.352139 & 5966 & 10.154 & yes \\
07821010 & 2938 & TYC 3146-1340-1& 291.3199 & 43.5955 & 24.238243  & 4969.615845 & 6298 & 10.816 & --- \\
08552540 & 7054 & V2277~Cyg      & 288.8904 & 44.6170 &  1.0619343 & 4954.105667 & 5749 & 10.292 & --- \\
09641031 & 7211 &FL Lyr, HD~179890&288.0203 & 46.3241 &  2.178154  & 4954.132713 & 5867 &  9.177 & --- \\
10031808 & 7278 & HD~188872      & 298.7976 & 46.9302 &  8.589644  & 4956.430326 &N/A$^c$& 9.557 & --- \\
10191056 & 5774 & WDS~J18555+4713& 283.8663 & 47.2283 &  2.4274949 & 4955.031469 & 6588 & 10.811 & yes \\
10987439 & 7396 & TYC 3561-922-1 & 296.8259 & 48.4434 & 10.6745992 & 4971.883920 & 6182 & 10.810 & --- \\
11922782 & 7495 & T-Cyg1-00246   & 296.0074 & 50.2326 &  3.512934  & 4956.247158 & 5581 & 10.460 & --- \\
\hline
\end{tabular}\\
$^a$ For the eclipsing binary, where $T_0$ is the primary eclipse mid-time;
$^b$ `yes' if lines from a third star are seen;\\
$^c$ No temperature given in the KEBC.
\end{table*}

For our program we selected targets from the \kep Eclipsing Binary Catalog
\citep[KEBC;][]{prs11,sla11,kir16}\footnote{\tt http://keplerebs.villanova.edu/}.
The basic target selection criteria were as follows:
\begin{enumerate}
\item \kep magnitude $kmag < 11$, to have the targets within the brightness range of 
the telescope.
\item Morphology parameter \citep{mat12} $morph < 0.6$ to exclude contact and semi-detached configurations.
\item Effective temperature from $T_{eff} < 6700$~K to have only late type systems, with many
spectral features. We querried the temperatures from the \kep Input Catalog 
\citep[KIC;][]{kep09}.
\end{enumerate}
So far in our program we have observed 21 objects, and published data for 
10 of them, and a publication dedicated to one more multiple system is in
preparation. In this work we present eight more system that are either double- 
or triple-lined spectroscopic binaries (SB2 or SB3). They are summarized in 
Table~\ref{tab_cat}. For each of them we briefly present the basic information below.
Unless stated otherwise, the eclipsing nature of a target was discovered by the
\kep mission, and no radial velocity data have been published till date. 
As all the KEBC eclipsing binaries, they all have their entries in the \kep Object of Interest 
(KOI)\footnote{\tt http://exoplanetarchive.ipac.caltech.edu/cgi-bin/TblView/nph-tblView?app=ExoTbls\&config=koi} 
database since DR24, and most of them are flagged as false positives.
The targets, obviously, appear in several catalogue papers related to the
\kep mission \citep[like][]{cou11,ten12,arm14}.

\begin{description}
\item {\it KIC~06525196 = KOI~5293, TYC~3143-604-1}:
It is a target with periodic eclipse timing variations (ETV; $P_{etv} = 415.8$~d), 
identified first by \citet{rap13}, and later by \citet{bor16}.
Both groups give the orbital parameters of the outer orbit.
Three narrow-line components are visible in the spectra, their RVs can be measured,
solutions for both orbits (inner -- eclipsing, and outer) can be obtained, and
masses of all three stars measured directly.

\item {\it KIC~07821010 = KOI~2938, TYC~3146-1340-1}:
This system has the longest orbital period in our sample, and a significant eccentricity. 
It probably hosts a non-transiting, circumbinary planet, which was first announced
in a conference presentation by 
W.~Welsh\footnote{\tt http://www.astro.up.pt/investigacao/conferencias/\\ 
/toe2014/files/wwelsh.pdf}
but the proper publication is still to be announced (Fabrycky et al., in prep.).
\citet{bor16} confirmed such possibility, by finding periodic modulations 
of the ETVs. Till now, absolute stellar parameters have also been presented on a conference
only, first by \citet{sha14}. Our observations started independently and simultaneously,
without prior knowledge of the work by Fabrycky~et~al., and our study utilizes only
our HIDES data.

\item {\it KIC~08552540 = KOI~7054, V2277~Cyg, T-Lyr1-00359, ASAS~J191534+4437.0, BD+44~3087}:
This star is classified in KOI database as a planetary candidate (PC) despite being
an eclipsing binary known before the \kep launch.
Discovered and first identified as a DEB by the Robotic Optical Transient Search 
Experiment~1 \citep[ROTSE1;][]{ake00}, first reported by \citet{die01}. 
Later also observed by the Trans-atlantic Exoplanet Survey \citep[TrES;][]{alo04}, 
and listed in the catalogue of variable stars in the \kep field of view of the 
All Sky Automated Survey \citep[ASAS-K;][]{pig09}.
By analysing the TrES light curve only, \citet{dev08} estimated the masses of both 
components: 1.655(15) and 1.296(13)~M$_\odot$ for the primary and secondary, 
respectively. Our spectroscopy allows us to revise these values. 
Several authors report ETVs \citep{gie12,gie15,con14},
but attribute them to the evolution of spots on the surface of both components.

\item {\it KIC~09641031 = KOI~7211, FL Lyr, HD~179890, HIP~94335}:
This is the only system in our sample with the full physical solution known before 
the \kep satellite was launched. Identified as eclipsing by \citet{mor35} 
and as (single-lined) spectroscopic by \citet{str50}. The most complete analysis 
so far was done by \citet{pop86}, who give 1.218(16)~M$_\odot$, 1.283(30)~R$_\odot$ 
for the primary, and 0.958(11)~M$_\odot$, 0.963(30)~R$_\odot$ for the secondary.
Recently, \citet{koz15} presented their own ETVs and claimed a discovery of a 
planetary-mass circumbinary companion candidate. They gave three possible solutions,
based on different orbital periods of the inner binary and the outer body. In
each case the time span of \kep data was shorter than the circumbinary period,
therefore all solutions are only preliminary and uncertain.

\item {\it KIC~10031808 = KOI~7278, HD~188872}:
The only star from our sample that has no $T_{eff}$ given in the KEBC, but 
a value of 6331~K can be found in the KOI database. Except brightness and 
position measurements, no literature data are available.

\item{\it KIC~10191056 = KOI~5774, WDS~J18555+4713, T-Lyr1-00687, ASAS~J185528+4713.7, BD+47~2717}: 
Another star classified in the KOI database as PC, despite 
being known to be an eclipsing binary. This is another triple-lined system in our sample. 
Reported by \citet{cou83} as a visual binary, with components of $V$ magnitudes 
10.97 and 14.80, separated in 1982 by 1.16 arcsec. Re-observed only recently by \citet{zie17},
who give the separation of 1.32 arcsec, and $\Delta mag=1.90\pm0.05$ at 6000~\AA\,
(from Robo-AO)~and $1.54\pm0.04$ in $K_p$ band (from Gemini-N/NIRI). Identified 
for the first time as an eclipsing binary by the TrES survey, and later
by ASAS-K. By analysing the TrES light curve only, \citet{dev08} estimated the masses of both 
components: 1.209(13) and 1.208(13)~M$_\odot$ for the primary and secondary, 
respectively. Our spectroscopy allows us to revise these values. In our spectra we 
see two sets of wider~lines, belonging to the components of the eclipsing pair, and 
another, narrow set coming from a third star. 
This system also has ETVs reported by several authors \citep{gie12,gie15,con14}, 
but no secure conclusions were drawn.

\item{\it KIC~10987439 = KOI~7396, TYC 3561-922-1}:
This system has the second longest period in our sample, and 
the narrowest spectral lines. Except brightness and
position measurements, no literature data are available.

\item {\it KIC~11922782 = KOI~7495, T-Cyg1-00246, TYC~3565-643-1}:
This system was first observed by the TrES survey
and identified as a detached eclipsing binary by \citet{dev08}. By analysing the light 
curve only they estimated the masses of both components: 1.498(26) and 
0.970(32)~M$_\odot$ for the primary and secondary, respectively. Our spectroscopy 
allows us to revise these values.
\end{description}

\section{Data and analysis}\label{sec_data}
Most of the methods and sources of data used in this work are identical to those
from Paper~I, and please refer to it for a detailed description. 
Here we only describe them briefly, and focus more on those 
that were not used in Paper~I due to a different nature of the researched objects.

\subsection{HIDES observations and RV measurements}
The spectroscopic observations were carried out during several runs between July 2014
and October 2016, at the 1.88-m telescope of the Okayama Astrophysical Observatory (OAO)
with the HIgh-Dispersion Echelle Spectrograph \citep[HIDES;][]{izu99}. The instrument 
was fed through a circular fibre, for which the light is collected via a circular 
aperture of projected on-sky diameter of 2.7 seconds of arc, drilled in a flat mirror 
that is used for guiding \citep{kam13}. An image slicer is used in order to reach 
both high resolution ($R\sim50000$) and good efficiency of the system. Spectra 
extraction was done under \textsc{iraf}, using procedures dedicated to HIDES.
Wavelength solution was based on ThAr exposures taken every 1-2 hours, which 
allows for stability of the order of $\sim$40~\ms. The resulting spectra span
from 4360 to 7535~\AA.

For the radial velocity measurements we used our own implementation of the 
\textsc{todcor} technique \citep{zuc94}, which finds velocities of two stars
$v_1$ and $v_2$ simultaneously. As templates we used synthetic spectra computed with 
ATLAS9 and ATLAS12 codes \citep{kur92}. Single measurement 
errors were calculated with a bootstrap approach \citep{hel12}, and used for 
weighting the measurements during the orbital fit, as they are sensitive to 
the signal-to-noise ratio ($S/N$) of the spectra and rotational broadening of the 
lines. 

The \todcor is optimized for double-lined spectroscopic binaries, but our sample 
includes triple-lined objects. For those, the velocities of the eclipsing pair 
were found from the global maximum, as each of these components contribute more 
to the total flux than the third star. The tertiary's velocities were found from 
a local maximum, where $v_1$ was for the tertiary, and $v_2$ for the primary of 
the eclipsing pair. 

In case of a hierarchical triple, for which we see motion of all components,
the outer orbits can also be modelled, or at least the mass ratio (tertiary 
to the inner binary) can be calculated (when the outer period is unknown). 
For this, we use the measured velocities of the third star, denoted as B, 
and variations of the systemic velocity of the inner pair A (Aa+Ab). 
In general, the two RVs of a spectroscopic binary -- $v_1,v_2$ -- 
measured at any time $t_i$, are related by:
\begin{equation}
q=\frac{v_1(t_i) - \gamma}{\gamma - v_2(t_i)}
\end{equation}\label{eq_q}
\citep{wil41} where $\gamma$ and $q$ are the systemic velocity and mass ratio 
(secondary over primary), respectively. Without a direct interaction between two 
stars, $q$ remains constant, but $\gamma$ can change in time, if the binary is 
orbited by another body. Assuming $q=const$ we can write:
\begin{equation}
\gamma(t_i) = \frac{v_1(t_i)+ q v_2(t_i)}{1+q}.
\end{equation}\label{eq_gam}
We used this formula to calculate $\gamma(t_i)$ of the Aa+Ab pair at a given moment
of observation $t_i$, and used these values in the orbital analysis of the A+B
outer orbit. The values of $q$ were taken from the solution of the inner
orbit.

All radial velocity measurements obtained from our HIDES spectra, together
with their errors and $S/N$ of the spectra, are listed in Table \ref{tab_RV} in 
the Appendix. For the triples KIC~06525196 and 10191056 we also
list the velocities of the third star, and measurements of $\gamma$ for each
observation (in the column `$v_1$').

\subsection{Publicly available data}
Other data used in this study are publicly available. 
The long cadence \kep photometry for all targets 
is available for download from the KEBC. It was the main source of the photometry 
used in this work. We used the de-trended relative flux measurements $f_{dtr}$, 
that were later transformed into magnitude difference $\Delta m=-2.5 \log(f_{dtr})$,
and finally the catalogue value of $k_{mag}$ was added. Short cadence data,
available for some of the targets, were not used due to their large amount and 
computer time required to analyse them.

Additional TrES data for KIC~08552540, 10191056, and 11922782 can be found in 
the on-line catalogue of eclipsing binaries identified by 
\citet{dev08}\footnote{\tt http://vizier.u-strasbg.fr/viz-bin/VizieR-3?-source=J/AJ/135/850/table7},
who used the LCs and information about total observed colours of a binary 
(but no spectroscopy) to directly determine the absolute masses and ages
of those systems. We include them in our work in order to directly
compare TrES and \kep curves, and our results with those from \citet{dev08}.

Two systems have $V$ and $I$-band LCs available from the ASAS-K 
website\footnote{\tt http://www.astrouw.edu.pl/asas/i\_kepler/kepler\_tab.html}.
These are KIC~08552540 = ASAS~J191534+4437.0 and KIC~10191056 = ASAS~J185528+4713.7.
The quality of data is rather poor, and the LCs do not contain many points, so the
eclipses are not always sampled properly. Nevertheless, we made an attempt to 
use them to asses the observed $V-I$ colours of each component, therefore
having independent estimates of temperatures.

Finally, we'd like to note that all systems have their LCs publicly available from the
SuperWASP archive\footnote{\tt http://exoplanetarchive.ipac.caltech.edu/cgi-bin/TblSearch/nph-tblSearchInit?app=ExoTbls\&config\\ =superwasptimeseries}. 
We have, however, decided not to analyse them. The photometric precision of
\kep is much better, these LCs were not used to determine absolute physical parameters,
and, themselves, they do not contain any important information 
that would not be possible to obtain from other data, like colours.

Archival RV measurements can be found for KIC~09641031 (FL~Lyr) in several 
literature sources, with the most recent RV curve published by \citet{pop86}. 
Before that, this system was
studied by \citet{str50}, but velocities of only one component were given.
We did not include the literature RVs in our study, and there is not much 
improvement when they are combined with our more accurate HIDES measurements.

\subsection{Eclipse timing variations}\label{sec_etv}
Four systems from our sample -- KIC~06525196, 07821010, 09641031, and 10191056 --
have been reported in at least one study to show eclipse timing variations (ETVs).
These are differences between the observed and predicted moments of eclipses,
which may be caused by a phenomenon intrinsic to the system (like spot evolution,
apsidal motion), or by a third body orbiting the eclipsing pair. 

Because not all authors publish their ETVs, and because the measured ETVs can
differ, depending on the method that was used, we decided to calculate our own.
As in Paper~I, we used the radio-pulsar-style method from \citet{koz11}. In this 
method, a template LC is created by fitting a trigonometric (harmonic) series 
to a complete set of photometric data. Then, the whole set of photometric data 
is divided to a number of subsets. Their number is arbitrary, but for this study
we set it to 200. For each subset, the phase/time shift is found by fitting 
the template curve with a least-squares method. In the final stage we removed
the obvious outliers, less than 10 in each case.  

We have calculated the ETVs for three systems: KIC~06525196, 09641031, and 10191056.
We omitted KIC~07821010 because of the ongoing analysis by Fabrycky et al. (in prep.).
Our ETVs are given in Table~\ref{tab_etv} in the Appendix.

\subsection{Orbital solutions}
The orbital solutions were found using our own procedure called {\sc v2fit}
\citep{kon10}. It is capable of working in different modes, and including
various optional effects (e.g. relativistic or tidal), but we used it mainly
in the simplest mode, where a double-Keplerian orbit is fit to a set
of RV measurements of two components, utilizing the Levenberg-Marquardt
minimization scheme. The fitted parameters are: orbital period $P$, zero-phase 
$T_P$\footnote{Defined as the moment of passing the pericentre for eccentric 
orbits or quadrature for circular.}, systemic velocity $\gamma$, velocity 
semi-amplitudes $K_{1,2}$, eccentricity $e$ and periastron longitude $\omega$, 
although in the final runs the last two parameters were usually kept fixed on 
values found by \jkt fit (Sect. \ref{sec_jkt}). Depending on the case,
we also included such effects as: the difference between systemic velocities of 
two components, $\gamma_2-\gamma_1$, linear and quadratic trends in $\gamma$, 
and periodic modulations of the inner binary'd $\gamma$, interpreted as 
influence of a circumbinary body on an outer orbit, parametrized analogously by 
orbital parameters $P_3$, $T_3$, $K_3$, $e_3$, and $\omega_3$. In such case 
$\gamma$ is defined in the code as the systemic velocity of the  whole triple. 
Whenever applicable, we simplified our fit by keeping the orbital period on the 
value given in the KEBC. Also, we were first letting $\gamma_2-\gamma_1$ to be 
fit for, but when the resulting value was indifferent from zero, we were keeping 
it fixed. 

Formal parameter errors of the fit are estimated by forcing the final reduced 
$\chi^2$ to be close to 1, either by multiplying them by a certain factor, or
adding in quadrature a systematic term (jitter). Because the code weights 
the measurements on the basis of their own errors, which are sensitive to 
$S/N$ and $v_{rot}$, we mainly used the first option in our analysis. The 
exception are active stars that clearly show spot-originated brightness 
variations. For these, the jitter compensates for the additional RV scatter
introduced by the spots. The errors given in Table \ref{tab_RV} are 
the multiplied ones, for which $\chi^2 \simeq 1$.

Systematics that come from fixing a certain parameter in the fit are assessed 
by a Monte-Carlo procedure, and other possible systematics (like coming from 
poor sampling, low number of measurements, pulsations, etc.) by a bootstrap 
analysis. All the uncertainties of orbital parameters given in this work 
already include the systematics.

\subsection{Light curve solutions and absolute parameters}\label{sec_jkt}

For the light curve analysis of \kep and TrES data we used version 28 (v28) 
of the code \jkt \citep{sou04a,sou04b}, which is based on the 
\textsc{ebop} program \citep{pop81}. As described in Paper~I,
the best fit was found on the complete Q0-Q17 \kep light curve,
but errors were estimated with a residual-shift (RS) method \citep{sou11},
run on data from each quarter separately. This approach is several
times faster than running RS on the complete curve, and properly includes
the influence of such systematic effects as spots or pulsations, which
we effectively treat as red (correlated) noise.

On the basis of spectroscopic data we first found the mass ratio $q$, 
as well as starting values of eccentricity $e$ and 
periastron longitude $\omega$. We fitted for the period $P$, mid-time of the 
primary (deeper) minimum $T_0$, sum of the fractional radii $r_1 + r_2$ 
(where $r = R/a$), their ratio $k$, inclination $i$, surface brightness 
ratio $J$, maximum brightness $S$, third light contribution $L_3$, as well 
as for $e$ and $\omega$ (their final values are from the \jkt runs,
unless stated otherwise). 

An initial model was done to roughly estimate the radii and masses
(to obtain $\log(g)$) and then the temperatures from the PAdova and 
TRieste Stellar Evolution Code 
\citep[PARSEC;][]{bre12}\footnote{\tt http://stev.oapd.inaf.it/cgi-bin/cmd}
isochrones. These results were used to estimate the limb darkening (LD) coefficients,
that were held fixed during the major fit, but perturbed during the RS stage.
We used the logarithmic limb darkening law \citep{kin70}, with the 
coefficients interpolated from the tables published on the PHOEBE
website\footnote{\tt http://phoebe-project.org/1.0/?q=node/110}. 
We found that results do not change significantly if we put LD coefficients
predicted for temperatures and gravities different by 150~K and 0.5~dex.
We also tried to set the LD free, but ended up in physically
impossible values. The gravity darkening coefficients were always kept 
fixed at the values appropriate for stars with convective envelopes ($g = 0.32$).

The results of LC and RV solutions were later combined in order to calculate 
the absolute values of stellar parameters using the \jktabs code, available 
together with the {\sc jktebop}. As an input, this simple procedure takes orbital 
period, eccentricity, fractional radii, velocity semi-amplitudes and inclination 
(all with uncertainties), and returns absolute values of masses and radii (in solar 
units), $\log(g)$ and rotational velocities, assuming tidal locking and 
synchronization. It can also calculate distance to an object, taking effective 
temperatures of two components, $E(B-V)$ and apparent magnitudes. The \jktabs 
does not work on brightnesses in \kep band, so, unless stated otherwise, for the 
distance estimation we used $B,V,J,H$ and $K$-band entries from {\it Simbad} 
\citep{wen00}. As the final value of distance we adopt a weighted average of 
five values, calculated for each band from the surface brightness-$T_{eff}$ 
relations of \citet{ker04}. For the majority of distance calculations, we used 
temperatures found from isochrone fitting (Sect.~\ref{sec_iso_met} and \ref{sec_age}). 
The ones found in KEBC correspond to the total system's light, and, especially in 
the case of triples, should not be taken as temperatures of the components. 

The problem is easier to solve when independent $T_{eff}$ estimates are known, 
like in the case of KIC~09461031 = FL~Lyr \citep{pop86}, or when multi-band 
light curves are available, so the components' individual colours can be estimated, 
like for the two systems with ASAS-K data. In these two cases, the \jkt fits were 
done to the $V$ and $I$-band data mainly to assess the fractional fluxes of each 
of the components, from which we inferred their observed magnitudes in both bands, 
and colours. Most of the parameters (like sum and ratio of radii, inclination, 
ephemerides) were therefore held fixed to values found for \kep data, and only 
surface brightness ratio, maximum brightness and third light were fitted for. 
The uncertainties were calculated with a Monte-Carlo procedure. 

We would also like to point out that, in principle, SuperWASP$-$\kep colours
are possible to obtain, but both filters are wide and cover similar
range of wavelength ($\sim$400--900 and $\sim$400--700~nm for \kep
and SuperWASP, respectively), therefore such colours would not carry much
information.

\subsection{Comparison with isochrones and age estimation}\label{sec_iso_met}

We use our mass and radius estimations to assess the age $\tau$ and evolutionary 
status of each system. We compare them with the PARSEC 
isochrones, which include values of absolute magnitudes in the \kep band. 
The isochrones were calculated for ages $\log(\tau$/yr) of 6.6 to 10.10, every 
$\Delta \log(\tau)=0.05$. In few cases the best match is obtained for $\tau$
not being a node of this grid of ages, like $\log(\tau)=9.72$. In such situations 
we generate a separate isochrone for the desired $\tau$.

For all systems we assume solar metallicity ($Z=0.0152$ in 
this set), because we lack $[Fe/H]$ or $[M/H]$ estimates that we find reliable. 
Those from, for example, KIC or MAST are calculated as for single stars, and usually 
base on photometry only. We do not always find an isochrone that 
agrees with our results at 1$\sigma$ level, but in no case the agreement is worse than 
3$\sigma$. In cases when two significantly different solutions are possible (i.e. 
resulting in main-sequence or pre-main-sequence stage), we also checked which one better 
reproduces the resulting flux ratio (from absolute magnitude difference). In all
such cases the main-sequence solution turned out to be the preferable one.

From the isochrones we determine the effective temperatures of components.
We take the $T_{eff}$'s from the isochrone that matches both stars best. For the
uncertainty, we take the difference between temperatures predicted 
by the isochrone matched to the whole system and the one that matches only the
particular component, rounded up to 50~K, which we believe is a reasonable 
precision. If the difference is small, i.e. $<100$~K, we assume 100~K as a 
conservative temperature error. We also give all $T_{eff}$'s with 50~K precision.
These temperatures
are later used for distance determination with \textsc{jktabsdim}. Because
the true information about the metallicity is missing, the resulting ages should 
be treated as preliminary, but the evolutionary stages should be reliable.

\begin{landscape}
\begin{table}
\centering
\caption{Orbital and physical parameters of eight double-lined eclipsing
binaries from our sample, obtained from \kep photometry and HIDES RVs.}\label{tab_par_sb2}
\begin{tabular}{lcccccccc}
\hline \hline
KIC			& 06525196~A$^a$& 07821010 	& 08552540$^b$ 	& 09641031	& 10031808 	&10191056~A$^{ab}$& 10987439 	& 11922782$^b$	\\
\hline
$P_{ecl}$ (d) 		& 3.42059774(14)&  24.238235(4)	& 1.06193441(4) & 2.17815425(7)	& 8.5896432(13) &2.427494881(19)&10.67459809(33)&  3.5129340(3) \\
$T_0$ (JD-2454900)$^c$	&  54.353595(29)&  69.61678(13)	&  54.105945(27)&  54.133349(3)	&  56.43099(10)	&  55.031699(5)	&  71.885044(32)&  56.24790(7)	\\
$T_P$ (JD-2454900)$^d$	&  53.508(12)	&  69.313(35)  	&  53.846(23)	&  53.5886(11) 	&  56.475(74)	&  53.931(18)	&  60.136(72)	&  51.856(20)	\\
$K_1$ (k\ms)		&  85.96(12)	&  66.64(34) 	& 121.0(1.6)	&  93.23(12) 	&  83.08(28) 	& 107.0(1.3)	&  76.39(10)	&  76.04(29)	\\
$K_2$ (k\ms)		&  91.62(22)	&  69.75(32) 	& 145.9(2.0)	& 118.19(30) 	&  80.42(15) 	& 119.3(1.0)	&  53.00(9)	&  97.01(42)	\\
$\gamma_1$ (k\ms)	&    (var)	& -17.18(9) 	& -14.1(1.1)	& -37.45(9) 	&  12.96(17) 	&    (var) 	& -19.11(11)	& -41.84(13)	\\
$\gamma_2-\gamma_1$ (k\ms)& 0.0(fix)	&   0.0(fix) 	&   0.0(fix)	&   0.0(fix) 	&  -0.17(21) 	&   0.0(fix) 	&  -0.37(19)	&   0.0(fix)	\\
$q$			&   0.9383(26)	&   0.9554(66)	&   0.829(16)	&   0.7888(22) 	&   1.033(4) 	&   0.897(13)	&   1.4413(31)	&   0.784(4)	\\
$M_1\sin^3(i)$ (M$_\odot$)& 1.0240(52) &   1.289(15)	&   1.144(36)	&   1.1923(66)	&   1.705(9)	&   1.536(31)	&   0.9776(34)	&   1.057(10)	\\
$M_2\sin^3(i)$ (M$_\odot$)& 0.9607(36) &   1.231(15)	&   0.948(28)	&   0.9405(35)	&   1.762(13)	&   1.378(35)	&   1.4090(45)	&   0.829(6)	\\
$a\sin(i)$ (R$_\odot$) &  12.010(17)	&  47.98(18)	&   5.604(54)	&   9.105(14)	&  26.723(54)	&  10.861(78)	&  27.272(28)	&  12.019(32)	\\
$e$			&   0.0(fix)	&   0.6791(14) 	&   0.0(fix)	&   0.0(fix) 	&   0.2717(14) 	&   0.00283(23)	&   0.0509(14)	&   0.0(fix)	\\
$\omega$($^\circ$) 	&    ---	&  58.86(13) 	&   ---		&   --- 	&  94.067(70) 	& 287.3(1.5)	&  51.7(2.3)	&   ---		\\
$r_1$			&   0.0874(85)	& 0.027852(114)	&   0.2509(31)	&   0.1389(25) 	&   0.09628(71)	&   0.1784(20)	&   0.03409(54)	&   0.1245(47)	\\
$r_2$			&   0.0877(85)	& 0.024383(57)	&   0.1806(40)	&   0.0995(26) 	&   0.11250(47)	&   0.1571(25)	&   0.05527(11)	&   0.0704(52)	\\
$i$ ($^\circ$)		&  85.15(34)	&  89.597(20)	&  85.83(46)	&  85.36(71) 	&  83.323(47) 	&  81.345(75)	&  85.614(66)	&  85.52(60)	\\
$J$			&   0.906(47)	&   0.857(21)	&   0.67(14)	&   0.434(49) 	&   0.903(34) 	&   0.9547(56)	&   2.56(22)	&   0.46(11)	\\
$L_2/L_1$		&   0.91(22)	&   0.726(13)	&   0.292(10)	&   0.224(35) 	&   1.2303(23) 	&   0.754(43)	&   6.48(1.14)	&   0.15(3)	\\
$L_3/L_{tot}$		&   0.243(9)	&   0.0(fix) 	&   0.0(fix)	&   0.0(fix) 	&   0.0(fix) 	&   0.1866(38)	&   0.0(fix)	&   0.0(fix)	\\
$M_1$ (M$_\odot$)	&   1.0351(55)	&   1.289(15)	&   1.153(36)	&   1.2041(76)	&   1.741(9)	&   1.590(32)	&   0.9862(34)	&   1.067(10)	\\
$M_2$ (M$_\odot$)	&   0.9712(39)	&   1.231(15)	&   0.956(28)	&   0.9498(46)	&   1.798(13)	&   1.427(36)	&   1.4215(45)	&   0.836(6)	\\
$R_1$ (R$_\odot$)	&   1.116(103)	&   1.3363(75)	&   1.410(22)	&   1.269(23) 	&   2.590(20)	&   1.960(26)	&   0.932(15)	&   1.501(57)	\\
$R_2$ (R$_\odot$)	&   1.057(103)	&   1.1698(52)	&   1.015(24)	&   0.908(24) 	&   3.027(14)	&   1.726(30)	&   1.512(31)	&   0.849(63)	\\
$a$   (R$_\odot$)	&  12.053(18)	&  47.98(18)	&   5.619(54)	&   9.135(17) 	&  26.905(54)	&  10.986(80)	&  27.352(29)	&  12.056(34)	\\
$\log(g_1)$	 	&   4.358(80)	&   4.297(4)	&   4.202(12)	&   4.312(16) 	&   3.852(6)	&   4.055(10)	&   4.493(14)	&   4.114(33)	\\
$\log(g_2)$	 	&   4.377(85)	&   4.392(3) 	&   4.406(20)	&   4.499(23) 	&   3.731(4)	&   4.119(15)	&   4.232(18)	&   4.503(64)	\\
$rms_{RV1}$ (k\ms)	&   0.26	&   0.15	&   3.02	&   0.42 	&   0.50	&   2.50	&   0.129	&   0.37	\\
$rms_{RV2}$ (k\ms)	&   0.41	&   0.48	&   5.26	&   0.71 	&   0.30	&   2.28	&   0.042	&   1.02	\\
$rms_{LC}$ (mmag) 	&   2.65	&   0.50	&  12.74	&   5.44 	&   1.22	&   0.41	&   0.072	&   8.06	\\
\hline 
\end{tabular}
\\$^a$ With RVs corrected for the motion around the common centre of mass with star B.
$^b$ From \jkt solutions for \kep curve only. 
\\$^c$ Mid-time of the primary (deeper) eclipse.
$^d$ Time of pericentre or quadrature.
\end{table}
\end{landscape}

\section{Results}\label{sec_results}

\subsection{Double-lined binaries}

In this Section we present the results of our analysis of eight
double-lined spectroscopic and eclipsing binaries. Their orbital and 
physical parameters are summarised in Table \ref{tab_par_sb2}. Presented 
uncertainties (1$\sigma$) include systematics, estimated with the 
MC+bootstrap (in \textsc{v2fit}) and residual-shift (in \textsc{v2fit}) 
methods. We follow the convention that the primary star is the one eclipsed 
during the deeper minimum, which for circular orbits means the hotter component.
The orbital period $P_{ecl}$ was taken from the initial \jkt run, and held 
fixed during the orbital fit. The model radial velocity and light curves are 
phase-folded with the period of eclipses found in \jkt and with zero-phase set to
the primary eclipse mid-time $T_0$.

\subsubsection{KIC~06525196~A}

\begin{figure*}
\includegraphics[width=0.9\columnwidth]{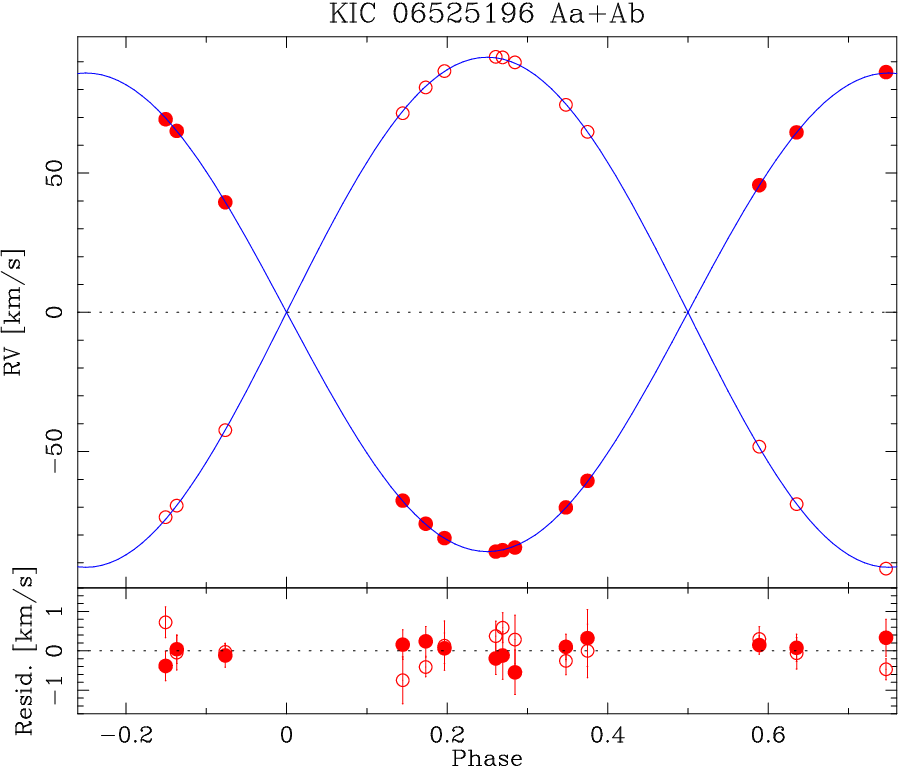}
\includegraphics[width=0.9\columnwidth]{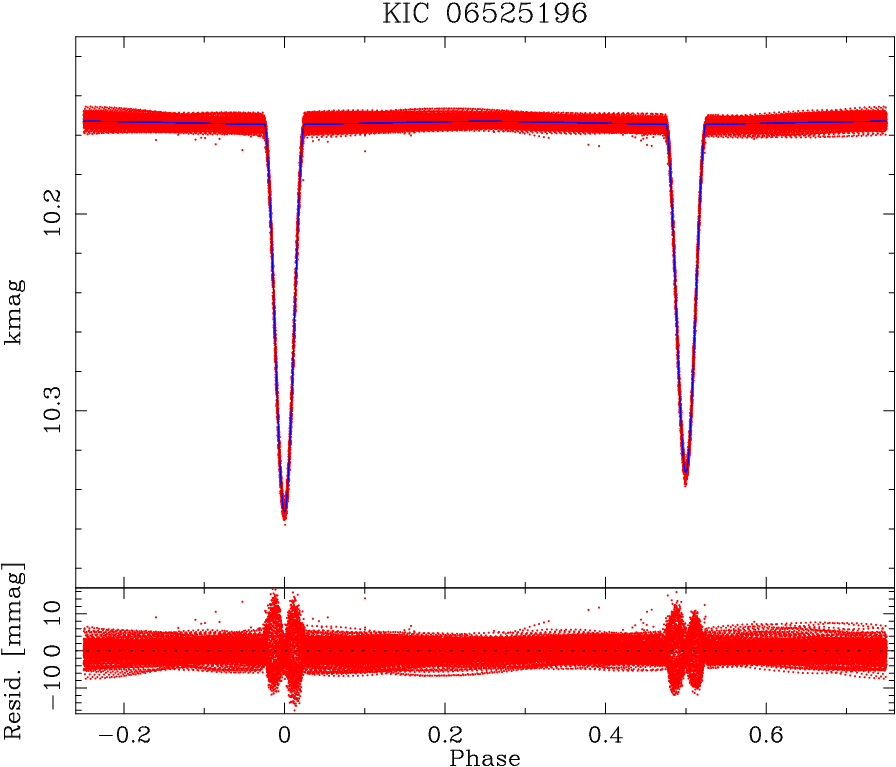}
\caption{Radial velocity (left) and light (right) curves of KIC~06525196. 
The RVs are for the components of the inner pair only, and are corrected for 
the motion on the outer orbit.
The best-fitting models are plotted with blue lines. Filled circles on the 
RV plot refer to the primary, and open ones to the secondary. The LC model
is fitted to the complete Q0-Q17 curve. Phase 0 is for the deeper
eclipse mid-time.
}\label{fig_mod_0652}
\end{figure*}

This is one of the two special cases of a triple-lined system in our sample, 
where three sets of lines are visible. In this hierarchical system the outer orbital
period is relatively short (418~d), and we covered it whole with our 
HIDES observations (time span of 676~d).
We use the convention that the inner pair is Aa+Ab, while the outer 
companion is B. Here we focus only on the Aa+Ab eclipsing pair, 
leaving the outer orbit of AB for discussion in a further Section.

We have observed this system 14 times with HIDES. Relatively narrow spectral
lines allow for quite precise RV measurements. In our orbital model we fitted
the parameter of the Aa+Ab orbit, and also for the periodic perturbation 
coming from the third star. We assumed that both the orbit of the inner pair
and the perturbation are Keplerian. We found no evidence for a non-zero 
eccentricity of the inner pair, nor for a difference in systemic velocities
of its components. We reached a very good precision of 0.5--0.4 per cent in 
masses, but a relatively poor one in radii: 9.2--9.7 per cent. This is 
mainly caused by the influence of the third body: its contribution to the
total flux, but also variations in the shape of eclipses of the phase-folded
light curve, coming from the fact that the moment of eclipses varies. This 
is clearly seen on the complete Q0-Q17 LC, shown in Figure~\ref{fig_mod_0652} --
note the shape of the residuals around eclipses (phases 0.0 and 0.5). One can
also note an out-of-eclipse variation, which we interpret as coming from 
spots that evolve in time. They also hamper the photometric solution.
In Fig.\ref{fig_mod_0652} we also show the RV curves of the inner binary, 
phase folded with its period and corrected for the influence of the third body.
The complete set of orbital and physical parameters is given in 
Table~\ref{tab_par_sb2}.

\subsubsection{KIC~07821010}

\begin{figure*}
\includegraphics[width=0.9\columnwidth]{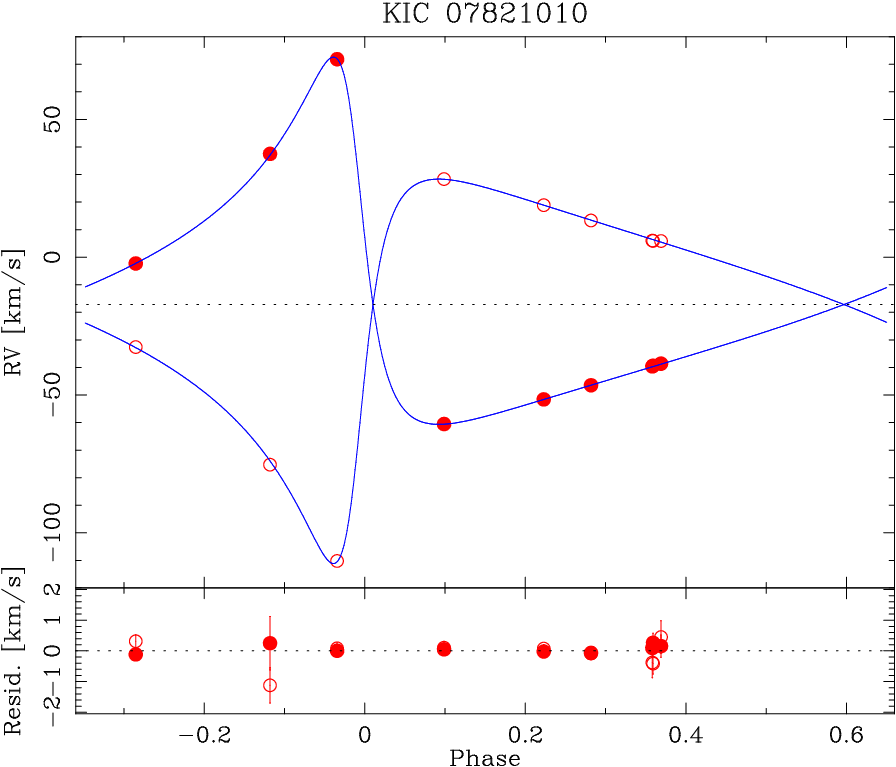}
\includegraphics[width=0.9\columnwidth]{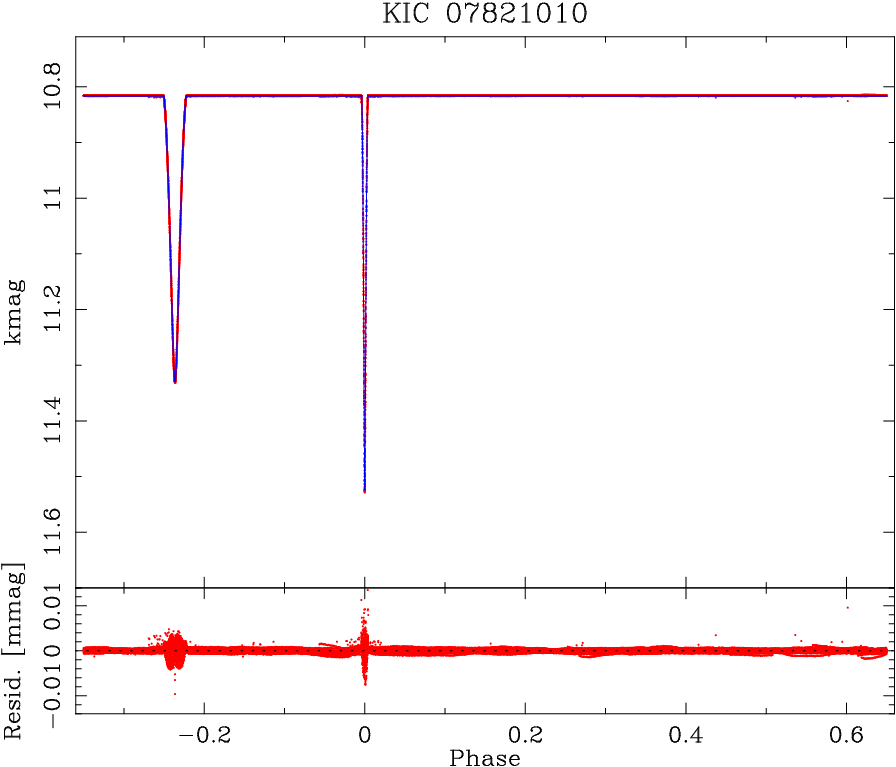}
\caption{Radial velocity (left) and light (right) curves of KIC~07821010. 
The best-fitting models are plotted with blue lines. Filled circles on the 
RV plot refer to the primary, and open ones to the secondary. The LC model
is fitted to the complete Q0-Q17 curve. Phase 0 is for the deeper
eclipse mid-time.
}\label{fig_mod_0782}
\end{figure*}

This star is the most eccentric and faintest SB2 in our sample, and the 
quality of HIDES data was highly dependent on weather conditions. Nine 
spectra were taken, but four of them are of lower $S/N$, which can be distinguished 
by the RV measurement errors, and they hamper the quality of the orbital fit. 
On the other hand the out-of-eclipse photometric variability is relatively low, 
so the LC-based parameters were found with very good precision. 

The observations and model RV and light curves are presented in Figure 
\ref{fig_mod_0782}. The strong deviations from the LC model seen around 
eclipses are possibly caused by a mismatch in LD coefficients. They also 
change from quarter to quarter, which can be explained by the presence of 
the circumbinary planet postulated by Fabrycky~et~al. As for other systems, 
we failed to find the LD coefficients when they were set as free parameters 
during the fit, but they were perturbed during the RS stage, so their 
influence on resulting parameters is accounted for (mainly on $e$, 
$\omega$ and $(R_1+R_2)/a$). In the end, we reached a good precision 
of $\sim$1.6 per cent in masses, and 0.7--0.9 per cent precision in radii. 
It is sufficient for reliable testing of evolutionary models \citep{las02}.

Our value of ratio of the radii $k=0.902(5)$ is not in a very good agreement 
with results from \citet{arm14}, who found $k=0.563(251)$. Their method bases 
on the total system's brightness measurements in several filters, and for this 
object produced large uncertainties. We thus concluded that their results 
should be treated with a lot of caution, and decided not to compare our 
results with theirs for other systems.

\subsubsection{KIC~08552540 (V2277~Cyg)}
\begin{figure*}
\includegraphics[width=0.9\columnwidth]{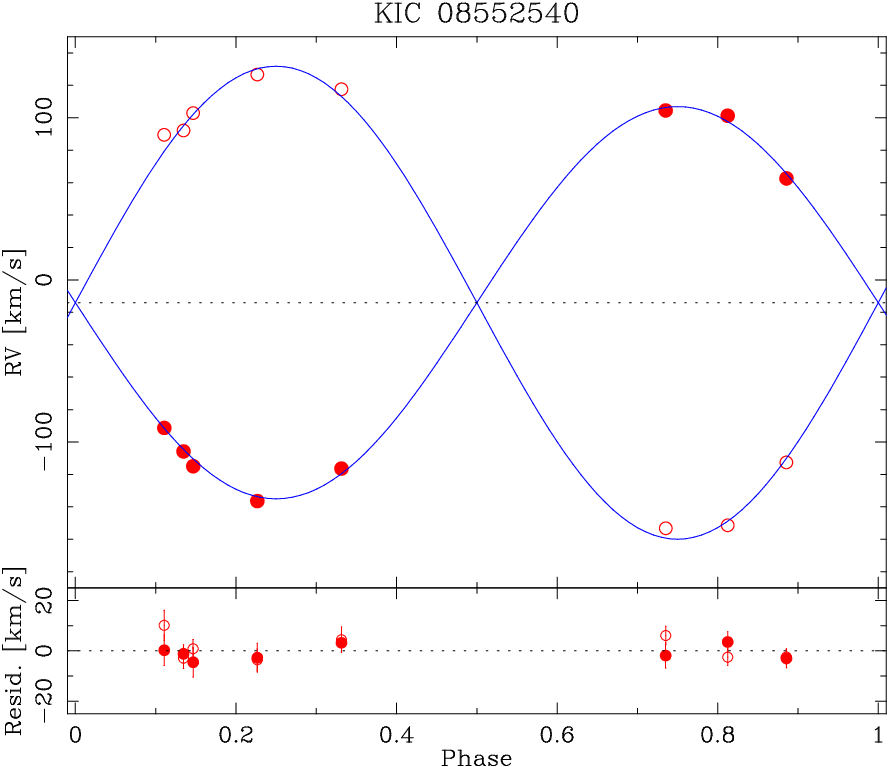}
\includegraphics[width=0.9\columnwidth]{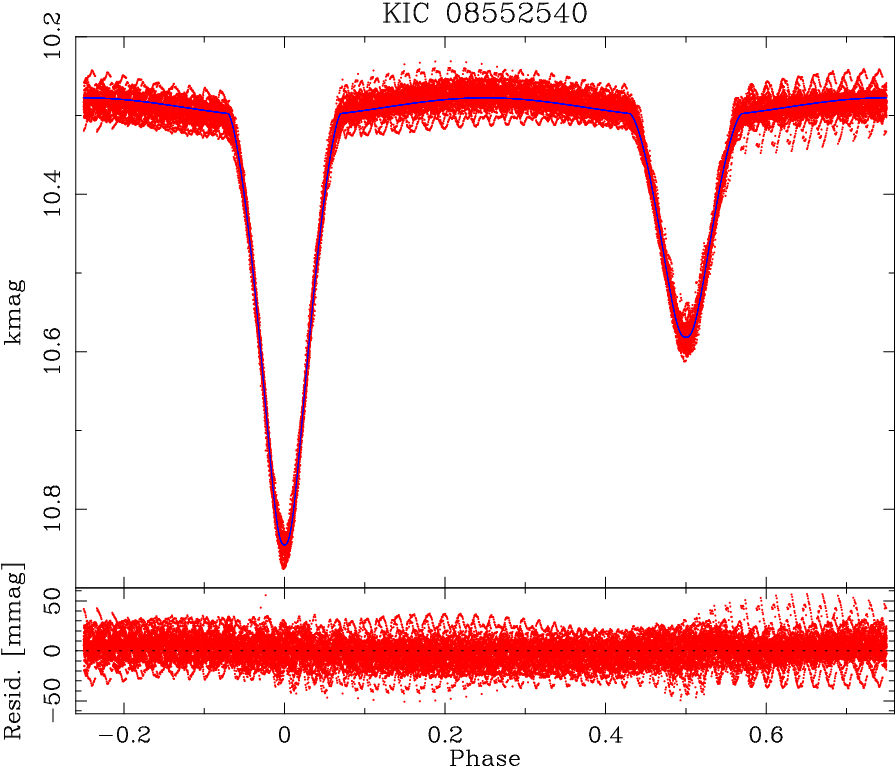}
\caption{Same as Fig. \ref{fig_mod_0782}, but for KIC~08552540.
}\label{fig_mod_0855}
\end{figure*}

This system has the shortest orbital period in our sample. In our eight 
HIDES spectra we see that the lines of both components are very broad, 
suggesting tidal locking and synchronous rotation. This clearly affected
the quality of the RV fit. Also, as seen in many late-type, short-period
binaries, there is a strong out-of-eclipse brightness modulation that
affected the LC modelling. From its character (sine-like shape, evolution in
time, variation of brightness in the minima) we conclude that it is caused 
by the presence of cold spots on both components.
Model curves and observations are shown in Fig.~\ref{fig_mod_0855}, and
parameters are listed in Table~\ref{tab_par_sb2}.
Despite large $rms$-es of both RV and light curves, the resulting 
uncertainties in masses and radii are quite low: 2.9-3.1 and 1.6-2.4 per 
cent, respectively.

It is worth to note that large spots on solar-mass components of short-period 
eclipsing binaries are not surprising, and were observed in other systems
\citep[e.g. CV~Boo: 1.032+0.968~M$_\odot$, $P=0.847$~d;][]{tor08}. 
They are a result of a presence of magnetic fields 
stronger than in single stars, enhanced by fast rotation in tidally-locked pairs.
Such a situation is common among lower-mass short-period systems.
In order to confirm the chromospheric activity, we examined 
the H$_\alpha$ lines in our spectra, but found no obvious emission features, 
although the $S/N$ is not always optimal, and the primary's absorption line may be 
partially filled. As explained in Paper~I, the Ca~II H and K lines are not in
the HIDES wavelength range.

\subsubsection{KIC~09641031 (FL~Lyr)}

\begin{figure*}
\includegraphics[width=0.9\columnwidth]{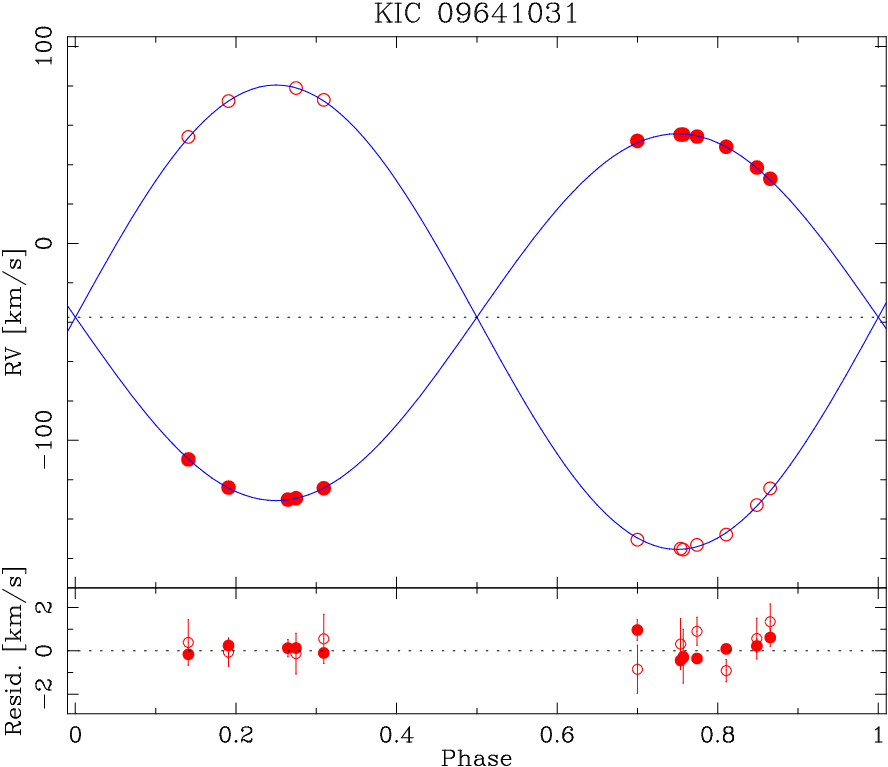}
\includegraphics[width=0.9\columnwidth]{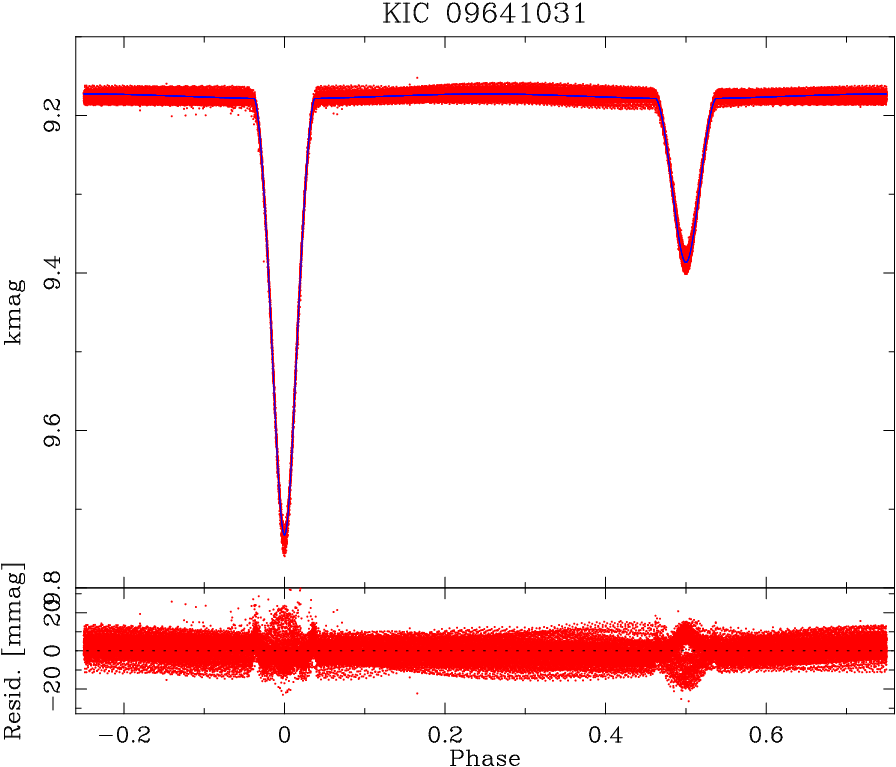}
\caption{Same as Fig. \ref{fig_mod_0782}, but for KIC~09641031.
}\label{fig_mod_0964}
\end{figure*}

This is the only system from our sample with RV measurements and full physical 
solution known before the \kep mission \citep{pop86}. It is also among the 
best-measured systems, that are listed in the on-line 
DEBCat\footnote{\tt http://www.astro.keele.ac.uk/$\sim$jkt/debcat/} catalogue 
\citep{sou15}. We have acquired twelve HIDES spectra, which is almost two times
fewer than in \citet{pop86}, but the RV precision is significantly better.
We have reached a very low uncertainty of 0.48-0.63 per cent in masses, which is
2-2.5 times better than previously. Our precision in radii is also good, and 
reaches 1.8-2.6 per cent, yet, surprisingly, it is only slightly improved in 
comparison with \citet{pop86}, despite superior photometric data. The explanation is, 
mainly, the influence of spots, clearly visible in \kep data, and slowly evolving 
in time. Our uncertainty in fractional radii comes in this case mainly from the spread of 
RS stage results for each separate quarter. Over the whole course of \kep observations 
the brightness modulation averages out in the LC, but for each quarter is slightly different.
It is worth to note that the spread of the \kep residuals is comparable to the spread
shown by \citeauthor{pop86}. 

\begin{table}
\centering
\caption{Comparison of our results for KIC~09641031 (FL Lyr) with parameters from 
\citet{pop86}.}\label{tab_fllyr}
\begin{tabular}{lcc}
\hline \hline
	&This work	& \citeauthor{pop86}\\
Parameter & (Table \ref{tab_par_sb2}) & (\citeyear{pop86})\\
\hline
$P$ (d)		& 2.17815425(7)	& 2.1781542(3)	\\
$K_1$ (k\ms)	&  93.23(12)	&  93.5(5)	\\
$K_2$ (k\ms)	& 118.19(30)	& 118.9(7)	\\
$r_1$		& 0.1389(25)	& 0.140(3)	\\
$r_2$		& 0.0995(27)	& 0.105(3)$^a$	\\
$i$ ($^\circ$)	& 85.36(71)	& 86.3(4)	\\
$M_1$ (M$_\odot$)	& 1.2041(76)	& 1.218(16)	\\
$M_2$ (M$_\odot$)	& 0.9498(46)	& 0.958(11)	\\
$R_1$ (M$_\odot$)	& 1.269(23)	& 1.283(30)	\\
$R_2$ (M$_\odot$)	& 0.908(24)	& 0.963(30)	\\
\hline
\end{tabular}
\\$^a$Calculated from the uncertainty of $R_2$ given in Table~17 of \citet{pop86},
which does not include all sources of errors. When calculated from the fractional
radius $r_1$, and ratio of radii $k$ (as in their Table 9), it becomes 0.007.
\end{table}

Our model is presented in Figure~\ref{fig_mod_0964}, with parameters in 
Table~\ref{tab_par_sb2}. They are compared with the solution by \citet{pop86} 
in Table~\ref{tab_fllyr}. One can see that we have improved the mass determination 
for this important system. Other parameters agree very well, with 
the exception of $r_2$ and $R_2$, for which our model gives values  
lower than in \citeauthor{pop86}, at the edge of 1$\sigma$ agreement. This
seemingly worse consistency is likely due to the fact that \citeauthor{pop86}
gave their uncertainty of absolute radii of the secondary underestimated (see the
comment under their Table~17). This is another reason why our results seem to be only 
slightly better. They also do not directly give the value (nor the error)
of the fractional secondary radius. In Table~9, they only give the fractional 
primary radius $r_p=0.140\pm0.003$, and the adopted ratio of radii $k = 0.75\pm0.05$.
When the fractional secondary radius and its uncertainty are calculated from these
values, with the proper error propagation, we obtain $0.105\pm0.007$. Our value
of $r_2=0.0995\pm0.0027$ is therefore well within their error.

\begin{figure}
\includegraphics[width=0.9\columnwidth]{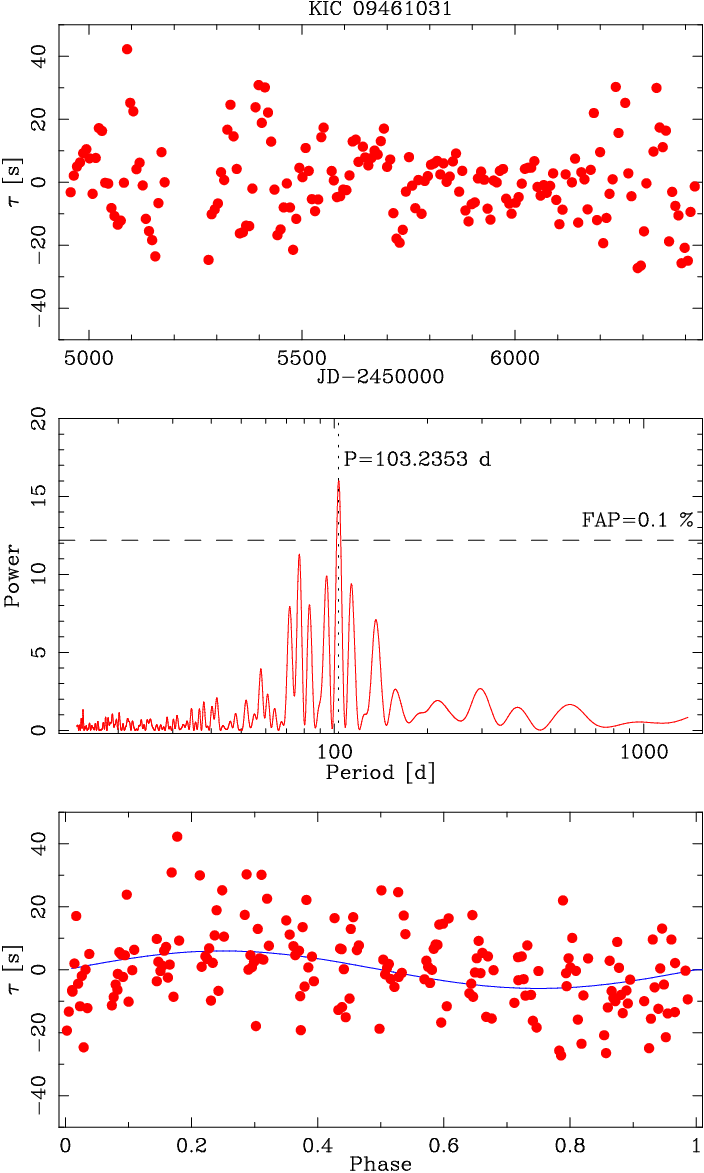}
\caption{{\it Top:} Our eclipse timing variation of KIC~09461031 as a function
of time. {\it Middle:} Their periodogram wit the tallest peak at $P=103.2$~d
marked. The horizontal line marks the false alarm probability (FAP) of
0.1 per cent. {\it Bottom:} The ETVs phase-folded with the resulting period.
The blue line is the best-fitting sine function. Its amplitude is $6.0\pm1.2$~s.
}\label{fig_etv_0964}
\end{figure}

Finally, we examine the claim of \citet{koz15} of detectable timing variations,
which they interpret as being caused by a third body on a long-period orbit.
In the Figure~\ref{fig_etv_0964} we present our own ETVs ($\tau$), 
calculated in a way described in Sect.~\ref{sec_etv}. We also show a 
Lomb-Scargle periodogram\footnote{Periodograms for this work were 
created with the on-line NASA Exoplanet Archive Periodogram Service:\\ 
{\tt http://exoplanetarchive.ipac.caltech.edu/cgi-bin/Pgram/nph-pgram}.}
of our measurements. The solutions from \citet{koz15} would produce a 
non-linear trend in the ETVs, which we don't see. In the periodogram, however,
we note a group of peaks, with the strongest one at $P=103.2$~d. Our ETVs
phase-folded with this period are also shown. We fitted a sine function to
them, and found the amplitude of $6.0\pm1.2$~s. After the fit, the $rms$ drops 
from 12.8 to 11.7~s.

The variation that we see seems to be statistically significant (FAP$<$0.1~per cent), 
however its origin remains unclear. If caused by a third body, its amplitude suggests
a low mass of the putative companion ($m \sin(i)\sim50$~M$_{JUP}$). Conversion from 
ETVs to RVs of the centre of the mass (modulation of the binary's systemic velocity; 
Paper~I) gives the RV amplitude of 1.26~k\ms. This is more than the $rms$ of the 
orbital fit (Tab.~\ref{tab_par_sb2}), so, in principle, we should be able to detect 
the signal. Unfortunately, our spectroscopic observations cluster around two phases 
of the putative outer orbit, when the predicted systemic velocities are similar, 
therefore we can not confirm the third-body scenario with our current HIDES data.
However, one should note that the LC shows a clear spot-originated modulation, and 
the observed ETVs might be a reflection of the evolution of spots. 
In any case, our results do not support the claim of a planetary-mass 
companion on a long-period orbit. 
 
\subsubsection{KIC~10031808}\label{res_1003}

\begin{figure*}
\includegraphics[width=0.9\columnwidth]{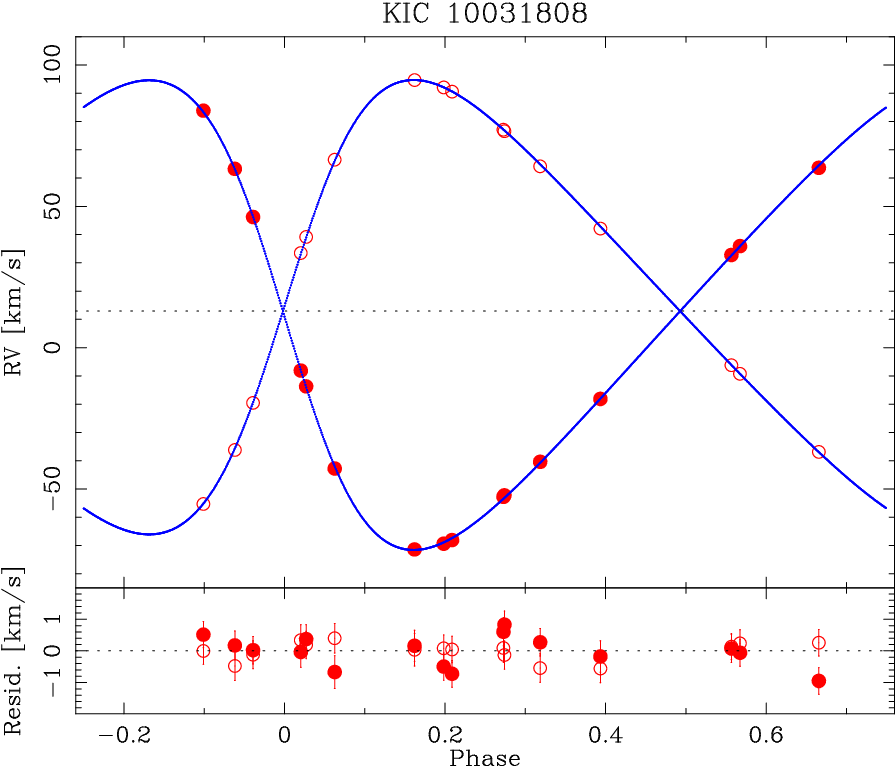}
\includegraphics[width=0.9\columnwidth]{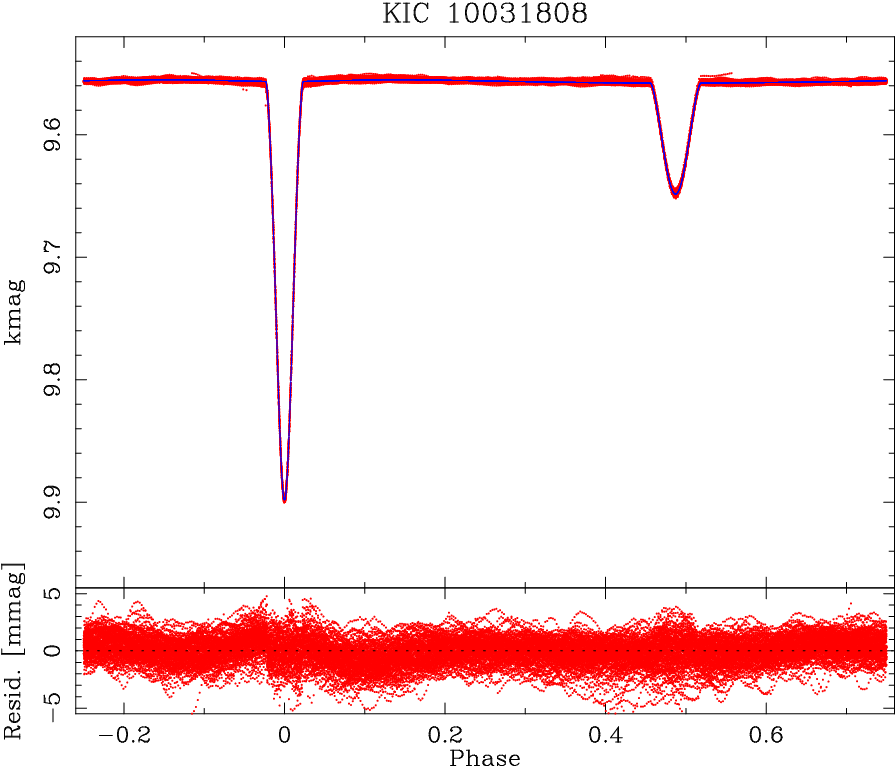}
\caption{Same as Fig. \ref{fig_mod_0782}, but for KIC~10031808.
}\label{fig_mod_1003}
\end{figure*}

This is the only star in our sample that does not have the temperature given in
the KEBC. We took 16 HIDES spectra of this system. Despite the two 
minima are separated in phase by nearly 0.5, a significant eccentricity was 
found in the LC modelling, that was nicely reproduced by the RVs. 
We had to fix $e$ when fitting each quarter light curve separately,
but we perturbed it during the RS stage. We also found small non-zero values of the
reflection coefficients: $1.88(8)\times10^{-4}$ and $1.33(7)\times10^{-4}$.

\begin{figure}
\includegraphics[width=0.9\columnwidth]{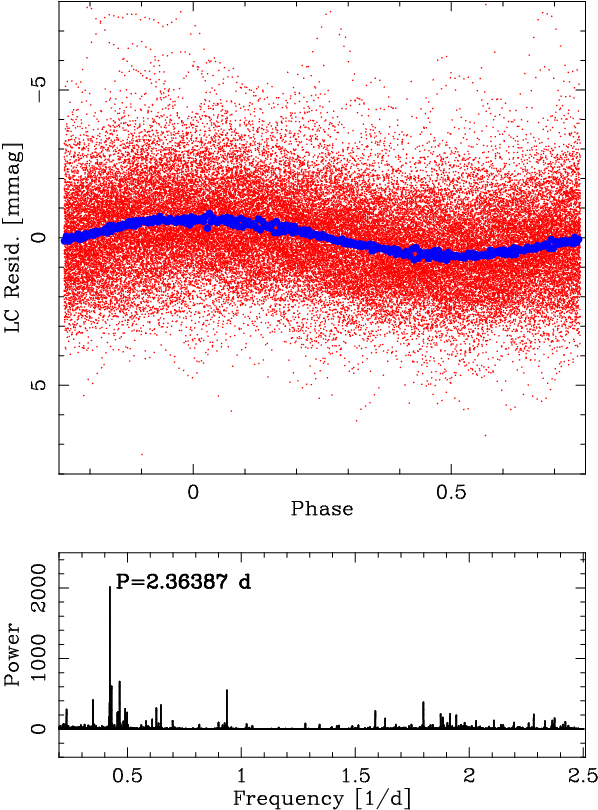}
\caption{Top: Residuals of the \jkt fit to the long-cadence LC of 
KIC~10031808 (red), phase folded with $P=2.36387$~d. Blue dots represent bins 
of 200 data points in the phase domain. Bottom: Piece of the periodogram of the
LC fit residuals, with he strongest peak at $P=2.36387$~d marked. Other
significant peaks are also seen, but none was found at frequencies $>$2.5~d$^{-1}$,
hence the cut-off.
}\label{fig_res1003}
\end{figure}

Results of the modelling are presented in Figure \ref{fig_mod_1003}, and 
parameter values can be found in Table \ref{tab_par_sb2}. We found that the
two stars are already evolved, currently at the end of the main sequence 
just before the transition to the giant branch. The more massive, larger 
component is cooler, and is the secondary in our nomenclature, because
it is eclipsed during the shallower eclipse. We reached a very good 
precision in both masses (0.5-0.72 per cent) and radii (0.46-0.77 per cent), 
which makes our results useful for testing the evolutionary models of the final 
stages of the main sequence. Precision in masses is slightly affected
by rotational broadening of the lines, but both components still seem
to rotate slower than synchronously (from {\sc jktabsdim}: 
$v_{synch}\simeq15$ and 18 k\ms \,for the primary and secondary, respectively).
Precision of radii (and other LC-based parameters) is also slightly 
hampered by additional photometric variability, although the $rms$ of the light curve
fit is relatively low. 

We run a Lomb-Scargle periodogram on the LC residuals, and found numerous
peaks at frequencies $<$2.5~d$^{-1}$, with the highest one corresponding 
to the period of 2.363871 days. The corresponding variability amplitude is about 
0.5~mmag (Fig. \ref{fig_res1003}). We attribute it to pulsations rather than rotation, 
as a rotation with such period would translate into velocity of 55 or 65 k\ms, 
depending on the component. The lines we observe in the spectra are not broadened that 
much. This still could be explained by a spin-orbit misalignment, but the 
results from \jktabs suggest alignment of orbital and rotational momenta
after $\sim$50~Myr. Meanwhile, as mentioned before, the system appears to be 
much older (see also Sect. \ref{sec_age}). The given period suggests
a $\gamma$~Doradus type of pulsations, and the lack of significant frequencies
higher than 2.5~d$^{-1}$ suggests no $\delta$~Scuti type variability.
We did not perform a detailed frequency analysis for this system, as it is
not the scope of this paper.

\subsubsection{KIC~10191056~A}

\begin{figure*}
\includegraphics[width=0.9\columnwidth]{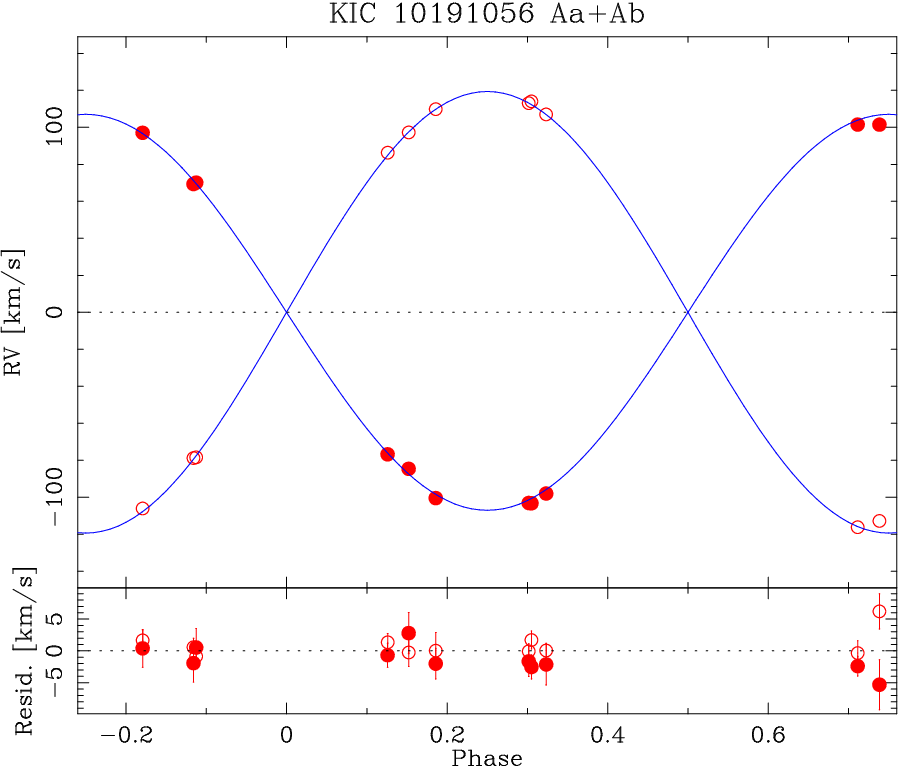}
\includegraphics[width=0.9\columnwidth]{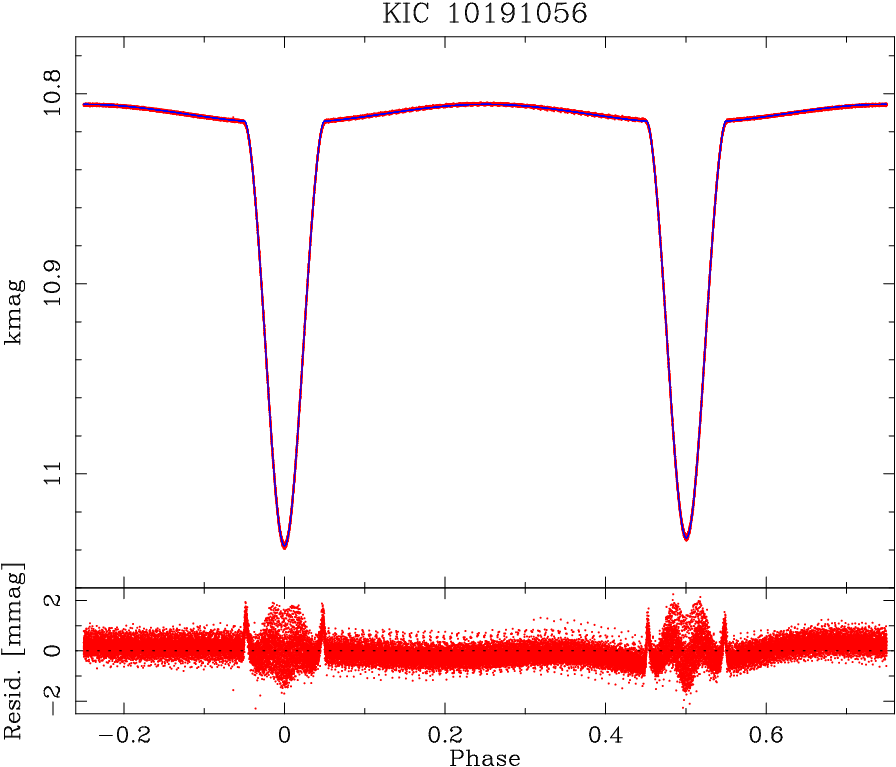}
\caption{Same as Fig. \ref{fig_mod_0652}, but for KIC~10191056.
}\label{fig_mod_1019}
\end{figure*}

Another object in this study that was detected previously by the TrES 
survey. It is also one of the two with three sets of lines visible in the spectra. 
As in KIC~06525196, also in this case the third star is fainter than the two components 
of the eclipsing pair. It has been observed with HIDES eleven times. The spectral 
lines of the eclipsing binary are rotationally broadened, which is expected from 
tidally-locked components of a $\sim$2.42-day pair. This, and low $S/N$ of
observations, hamper the precision of the RV-based parameters.

The \kep light curve, on the other hand, is only weakly affected by variability of
a kind other than eclipses and ellipsoidal variations, like spots or oscillations.
Their amplitude is only about 1~mmag. It is in agreement with the fact that 
KIC~10191056 has the highest value of $T_{eff}$ listed in the KEBC among our targets. 
The LC is more affected by some other systematic 
effects, like incorrect de-trending. For this reason we removed from the LC
the data from quarters 14-17, and small pieces from quarters 2 and 10, still
leaving almost 44000 data points from quarters 0 to 11 (there are no long-cadence
data from quarters 12 and 13).

The careful analysis of the LC also revealed that the orbit is actually not circular.
There is a small displacement of the secondary eclipse from the exact phase 0.5, 
by $\sim$18.7~min, or 0.00053~$P$. The measurements of eclipse times, given
separately for the primary and secondary by \citet{gie15}, seem to confirm that
by showing a gradual diverging, which suggests apsidal motion. The RV fit was 
therefore done with values of $e$ and $\omega$ fixed to those found in the LC fit. 
The ETVs of \citet{gie15} also show a small curvature (non-zero time derivative of 
the orbital period, $\dot{P}$), which can be explained either by a presence of a third 
body, or a mass transfer. The latter seems unlikely, as the stars are far from
filling their Roche lobes, and a clear detection of a third set of lines in the spectra 
supports the former scenario. The model presented in Figure \ref{fig_mod_1019}, with 
parameters listed in Table \ref{tab_par_sb2}, has been prepared under this assumption, 
i.e. the third light and linear variation of the systemic velocity have been 
accounted for. See Section \ref{sec_K1019_B} for the discussion of the 
motion of the companion.

We have reached a satisfactory precision of 2.0--2.5 per cent in masses, and
a slightly better level of 1.3--1.7 per cent in radii, hampered mainly be the uncertainty
in the third light. The uncertainty in mass already takes into account the error
in the linear trend, which is discussed in Section \ref{sec_K1019_B}.

\subsubsection{KIC~10987439}

\begin{figure*}
\includegraphics[width=0.9\columnwidth]{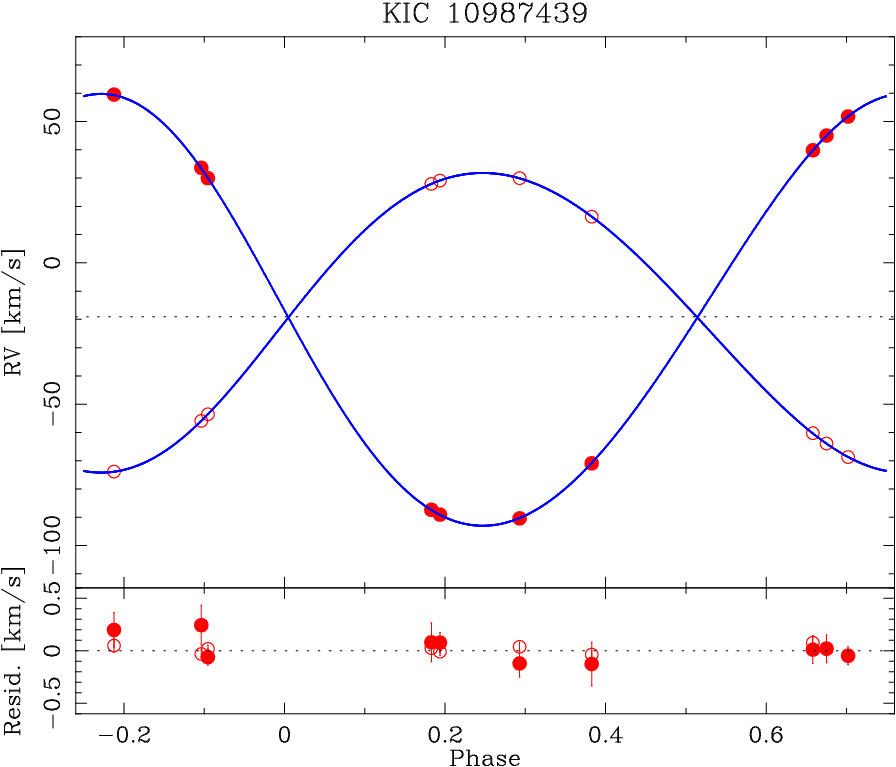}
\includegraphics[width=0.9\columnwidth]{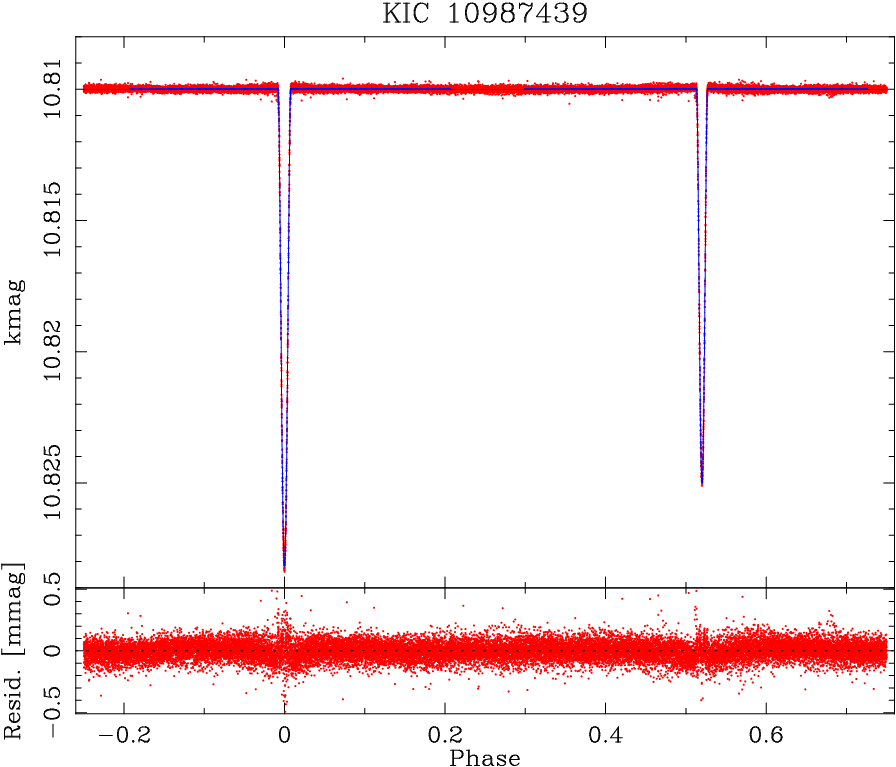}
\caption{Same as Fig. \ref{fig_mod_0782}, but for KIC~10987439.
}\label{fig_mod_1098}
\end{figure*}

This system has the lowest amplitude of the out-of-eclipse variability in our sample,
resulting in the smallest $rms$ of the LC model 0.072 mmag ($\sim66$~ppm in flux).
Also, the amount of photometric data is the lowest -- only $\sim$25000
long-cadence measurements from quarters 1, 5, 9, 10, 13, 14, and 17, and no
short-cadence data at all. 

\begin{figure}
\includegraphics[width=0.9\columnwidth]{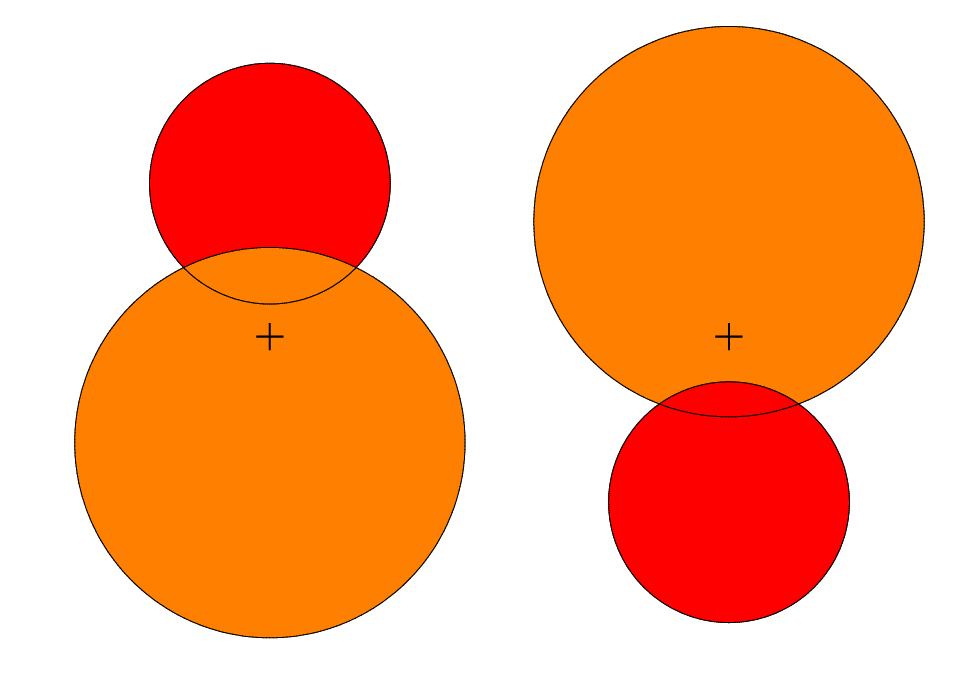}
\caption{Simplified illustration of the eclipses in KIC~10987439. In the primary eclipse
(left) the smaller, cooler component (red circle) is behind the hotter, larger one (orange).
In the secondary eclipse (right) the situation is opposite. The eclipse of the cooler star 
is deeper because larger area is obscured. Black contours are plotted to help to compare the 
two obscured areas by eye. Black crosses mark the centre of mass.
}\label{ecl_1098}
\end{figure}

Despite similar depth of the eclipses, the pair turned out to be composed of quite 
different stars, and, to our initial surprise, it is the cooler, smaller, and less 
massive component that is eclipsed during the primary (deeper) minimum. This is because 
of a projection effect, caused by a small, but measurable eccentricity, and inclination
relatively far from 90 degrees (see Fig. \ref{ecl_1098}). The eclipses are 
shallow and only grazing, the primary one lasts about 20 per cent longer, and larger 
part of the stellar disk is obscured.

The spectroscopic data are also of a very good quality. The star has been observed 
ten times with HIDES. Narrow spectral lines made it possible to achieve the best 
RV precision in our sample for a single component, with the $rms$ of 42~\ms 
\,for the component that contributes more to the system's total flux (86.5 per cent), 
but is the secondary according to our convention. This is the level of RV precision
we reached for RV standards (see Paper~I). We also reach quite a good $rms$ of 
129~\ms for our faint primary.

The light and RV curves are shown in Figure~\ref{fig_mod_1098}, and the system's
parameters can be found in Table \ref{tab_par_sb2}. As in the case of KIC~10031808,
we had to fit for reflection coefficients, and found them to be $3.42(2)\times10^{-4}$
and $3.41(2)\times10^{-4}$ for the primary and secondary, respectively. We reached a 
very good precision in masses: $\sim$0.33 per cent for both components, but a 
significantly worse in radii: 1.6--2.0 per cent. This seems surprising considering
the excellent $rms$ of the LC fit, but it may be understood when one takes the
grazing eclipses into account. In such situations, there is not only a strong 
degeneration of the orbital inclination $i$ with the ratio of the radii $k$, but also
with their sum ($r_1+r_2$). A small change in $i$ leads to a change in $r_1+r_2$
that is relatively large, in comparison to a situation when the eclipses are nearly central.
Still, our results are good enough for meaningful tests of stellar evolution models.

There is also some leftover variability in the LC's residuals. Larger scatter around
the primary minimum suggests presence of small, migrating spots on the surface of the 
primary, whose spectral type is probably late G or early K. Due to the lack of
short-cadence data, the spots can not be properly monitored, therefore we have removed 
about 20 most deviating data points from the minima, as they were causing the model 
to underestimate the eclipses' depths, and hampering the results. Outside of the 
eclipses there is also a periodic modulation of an amplitude of 0.032~mmag 
present, and probably weak flares. The latter, if present, were however 
short-lasting, and, with the $\sim$30~min data cadence, covered by only 1-2 data points.
The periodic modulation may be produced by spots, but with period of 1.624 days this would
imply an asynchronous rotation with velocities of the order of 20~k\ms. We do not 
observe lines broadened that much -- the stars rather seem to rotate synchronously. 
The planes of rotation may still be different than
the orbital plane, but the orbit itself is nearly-circular, and, according to the
theory of tidal interactions, the circularisation of the orbit occurs much later
($\sim10^{11}$~yr) than synchronisation of orbital and rotational periods, 
and spin-orbit alignment ($\sim10^8$~yr). 

\subsubsection{KIC~11922782}

\begin{figure*}
\includegraphics[width=0.9\columnwidth]{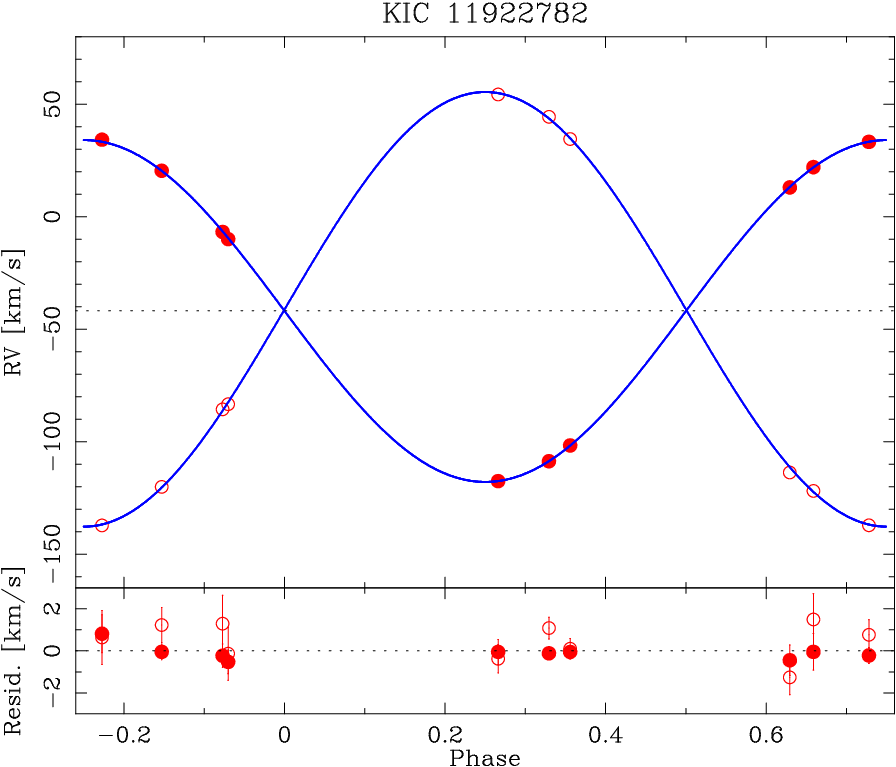}
\includegraphics[width=0.9\columnwidth]{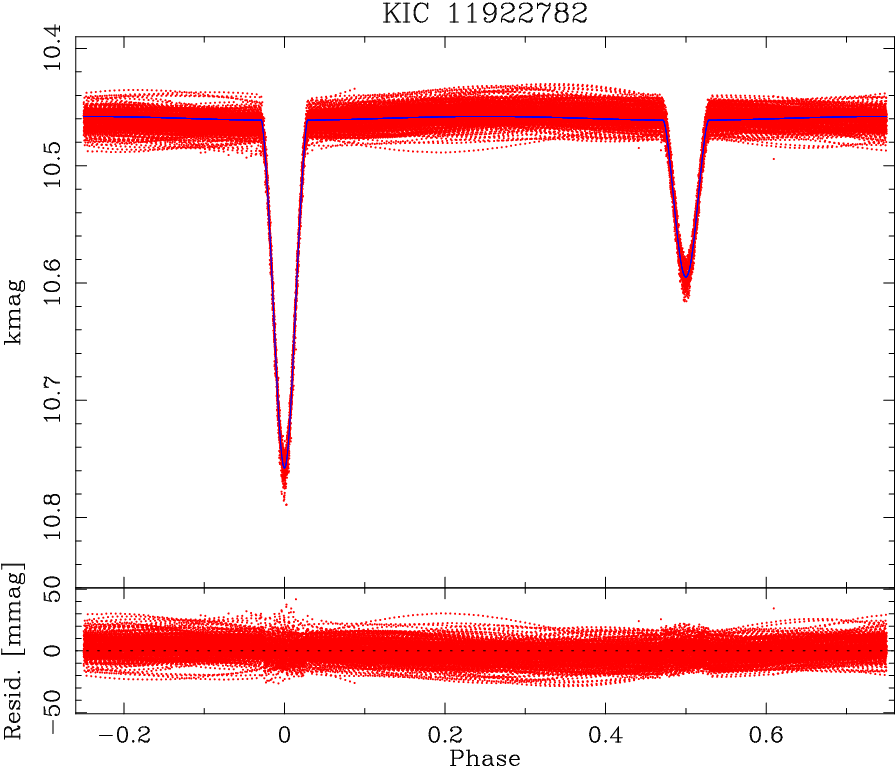}
\caption{Same as Fig. \ref{fig_mod_0782}, but for KIC~11922782.
}\label{fig_mod_1192}
\end{figure*}

The last system in our sample, and the third common with \citet{dev08},
was observed with HIDES ten times. The model we obtained for this pair is 
presented in Figure \ref{fig_mod_1192}, with parameters listed in 
Table~\ref{tab_par_sb2}. We have reached a very good precision in masses 
(0.7-0.9 per cent), but significantly worse in radii (3.8-7.4), hampered 
mainly by spots evolving in time (note the large $rms$ of the LC fit). 
This system also has one of the lowest mass ratios, and the highest contrast 
(in terms of luminosity ratio) between the components, which is reflected 
by very different $rms$-es in the RV fit, and contributed to large uncertainties 
of the LC-based parameters. The secondary is also the lowest-mass star we have 
analysed in this paper, (0.835~M$_\odot$) and discrepancies between the
observed and theoretically predicted radius are expected. 
The primary's mass is very close to solar, but its radius 
is much larger, therefore we probably deal with an evolved version of our Sun.
It is therefore an interesting system for further studies.

The light curve is strongly affected by rapidly evolving spots, located
on both components, which can be deduced from variations in the depth
of both minima. In order to confirm the chromospheric activity, we examined 
the H$_\alpha$ lines in our spectra, but found no obvious emission features, 
although the $S/N$ is not optimal, and the primary's absorption line may be 
partially filled. As mentioned before, the Ca~II H and K lines are not in
the HIDES wavelength range.

\subsection{Tertiary components}

In this section we focus on the motion of the outer components of two 
triples: KIC~06525196 and 10191056. We follow the convention
that the outer star is designated as the component B, and the inner eclipsing 
binary as A (= Aa+Ab). 

\subsubsection{The outer orbit of KIC~06525196}\label{sec_K0652_B}

\begin{table}
\centering
\scriptsize
\caption{Parameters of the outer orbit and the tertiary star of KIC~06525196
obtained in this study and compared with the literature.}\label{tab_par_3rd}
\begin{tabular}{lccc}
\hline \hline
	 & This & \citeauthor{rap13} & \citeauthor{bor16} \\
Parameter& work & (\citeyear{rap13}) & (\citeyear{bor16}) \\
\hline
$P_{AB}$~(d)		& 418.0(4)	& 415.8(-) & 418.2(1) \\
$T_{per}$ (JD-2450000)	& 6746.7(9)	& 6805$^{+13}_{-60}$& 6743(3) \\
$K_A$ (k\ms)		& 11.67(12)	& ---	& --- \\
$K_B$ (k\ms)		& 30.15(17)	& ---	& --- \\
$q_{AB}$		& 0.3872(45)	& 0.61$^{+0.77}_{-0.24}$& 0.41(6)\\
$e_{AB}$		& 0.301(3)	& 0.30$^{+0.05}_{-0.04}$& 0.295(5) \\
$\omega_A$ ($^\circ$)	& 276(1)	& 285$^{+45}_{-52}$	& 274(2)\\
$\gamma$ (k\ms)	& 4.58(4)	& ---	& --- \\
$M_A \sin^3(i_{AB})$ (M$_\odot$)& 1.981(30)& 0.85$^{+2.60}_{-0.69}$ & --- \\
$M_B \sin^3(i_{AB})$ (M$_\odot$)& 0.767(14)& 0.59$^{+0.74}_{-0.44}$ & --- \\
$a_{AB}\sin(i_{AB})$ (AU)	 & 1.532(8) & --- & ---\\
$i_{AB}$ ($^\circ$)	& 84.7$^{+5.3}_{-2.6}$	& --- & 80(-)\\
$M_A$ (M$_\odot$)	& 2.0063(67)$^a$& ---	& 2.0(5) \\
$M_B$ (M$_\odot$)	& 0.777(12)$^b$	& ---	& 0.8(2)\\
$a_{AB}$ (AU)	& 1.539(10)& --- & 1.55(13) \\
$rms_A$ (k\ms)	& 0.182 & ---	& --- \\
$rms_B$ (k\ms)	& 0.096 & ---	& --- \\
\hline
\end{tabular}
\\$^a$ Directly from Table \ref{tab_par_sb2}.
$^b$ From $M_A$ and $q_{AB}$
\end{table}

\begin{figure}
\centering
\includegraphics[width=0.9\columnwidth]{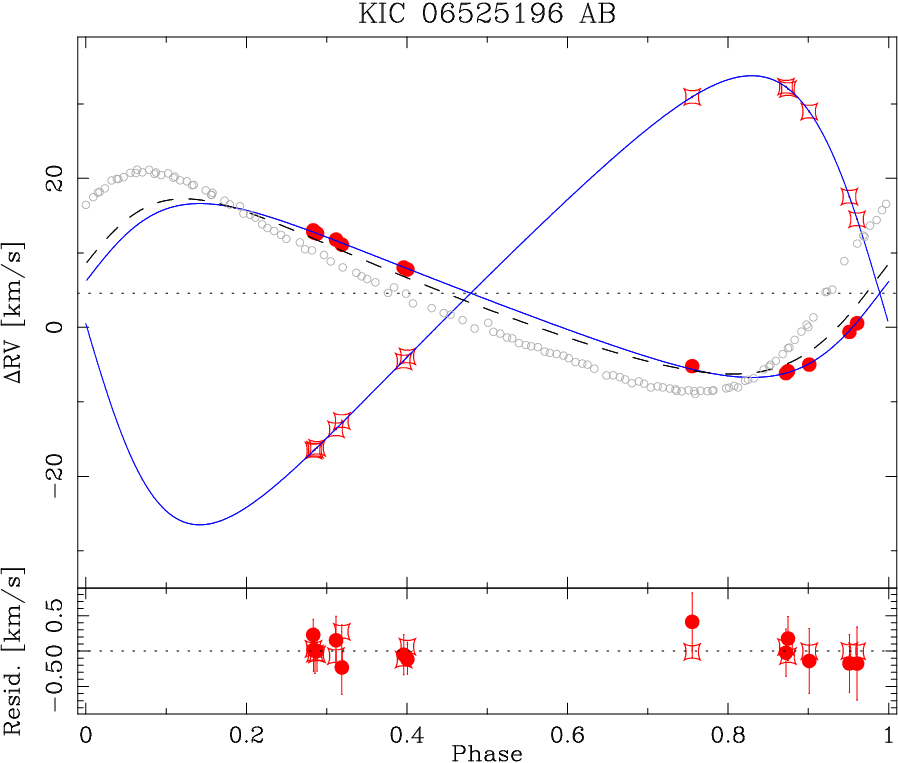}
\caption{Solution for the outer orbit of the KIC~06525196
triple-lined system. Filled red dots show the calculated systemic velocity
of the inner pair Aa+Ab, open red symbols denote the direct RV measurements 
of the third star. Grey dots represent RVs obtained from directly translating
the ETVs. They clearly show larger amplitude and phase shift, because the 
dynamical timing variation effect has not been removed. The black dashed line 
is the RV curve predicted from the solution of \citet{rap13}, but taking only 
the R\"{o}mer amplitude of the ETVs. The dotted line marks the systemic velocity of 
the whole triple.
}\label{fig_orb3_0652}
\end{figure}

\begin{figure}
\includegraphics[width=0.9\columnwidth]{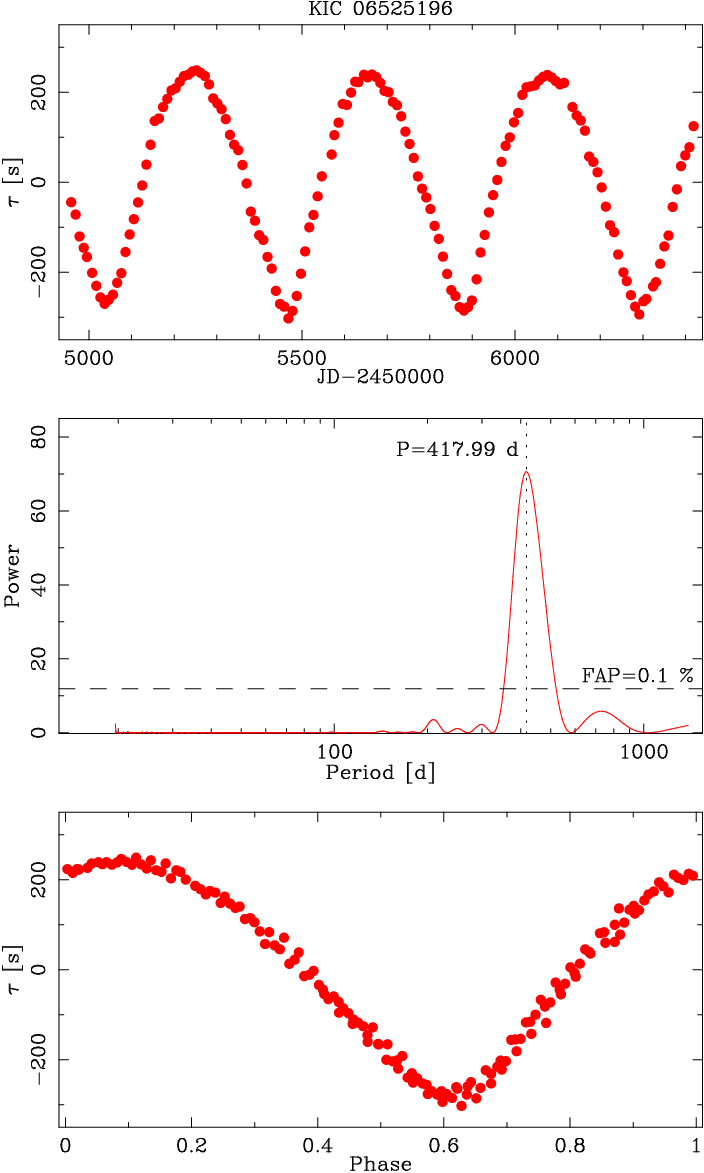}
\caption{Same as Fig.~\ref{fig_etv_0964}, but for KIC~06525196. 
The period of the outer orbit (418~d) is clearly seen on the periodogram.
}\label{fig_etv_0652}
\end{figure}

This is a case of a triple-lined spectroscopic system, for which
velocities of all stars were measured. The parameters of the outer orbit
were previously estimated from ETVs \citep{rap13,bor16}, and the period 
is short enough to be covered with observations during only few semesters.
We assumed that the outer orbit is Keplerian, as the distance to the third star 
B is much larger than the separation between Aa and Ab. In other words, we treated 
the outer A+B pair as a binary, but as the RVs of A we used the Aa+Ab systemic 
velocity measurements, calculated using the Equation~\ref{eq_gam}.
These measurements are also listed in Tab.~\ref{tab_RV}.

The results are summarised in Table \ref{tab_par_3rd}, and presented
in Figure~\ref{fig_orb3_0652}. As previously, 
uncertainties include systematics calculated with a bootstrap routine. 
Among other parameters, we show times of the pericentre passage $T_{per}$, 
and the pericentre longitude given for the centre of mass of the inner binary. 
We also give the systemic velocity of the whole triple as $\gamma$.
Because absolute mass of A is known (Tab.~\ref{tab_par_sb2}), we
could calculate the inclination of the outer orbit, and found that the
orientation is nearly edge-on, and $i=90^\circ$ is possible. There are 
however no signs of tertiary eclipses, nor eclipse depth variations,
suggesting large mutual inclination $i_m$. Such a scenario has been found 
by \citet{rap13}, who give $i_m$ between 22.6 and 33.9$^\circ$, but other 
parameters show that their solution is only in a marginal agreement with ours.
Moreover, their analysis of the LC led to very different values of the mass
ratio of the inner binary, and third-light contribution: 0.71(1) and 0.024(1),
respectively. 

In much better agreement with ours are the parameters given by \citet{bor16}.
They, however, assume the mutual inclination to be zero, and argue that a small 
non-zero value ($i_m<10^\circ$) would produce significant eclipse depth variations, 
and that their solution rules out large values of $i_m$. Table \ref{tab_par_3rd} 
shows a comparison of our results with those obtained by \citet{rap13} and \citet{bor16}.

In the Paper~I we have introduced a method to translate the observed ETVs
to RVs of the centre of mass of the eclipsing binary. With the solution of the
outer orbit known quite well, KIC~06525196 appears to be a good example to test
this method of translation. We calculated our own ETVs, using the method described in
the Section~\ref{sec_etv}. We show them in Figure~\ref{fig_etv_0652}. They are 
consistent with those from \citet{rap13} and \citet{bor16} in terms of the orbital 
period (418~d) and total amplitude ($\sim$250~s). However, by translating them 
directly to the RVs, we get a result that is not consistent with what we have measured 
(Fig.~\ref{fig_orb3_0652}). The amplitude is larger ($\sim$13~k\ms), and there seems 
to be a shift in phase. 

It is because the ETVs in this system are caused by two effects -- the 
`classical' light time travel effect (LTTE, a.k.a. the R\"{o}mer delay), 
and dynamical perturbations from the third body. \citet{rap13} clearly distinguish 
the contribution of those two effects, and in their Figure~2 one can see that 
their maxima are shifted in the orbital phase. They also give values of 
amplitudes of the two effects separately: 215~s for the LTTE, and 127~s 
for the dynamical (\citeauthor{bor16} only give their ratio). Using the 
Equation~8 from Paper~I, we can estimate the RV amplitude expected from 
the LTTE to be 11.76~k\ms, which is very close to what we have actually given 
in Table~\ref{tab_par_3rd}. The corresponding RV curve is also drawn in 
Fig.~\ref{fig_orb3_0652} as the black dashed line. It is similar to our solution, 
with the main discrepancy coming probably from a different period and moment of 
periastron passage. This case clearly shows that the RV motion of the centre of 
mass (variations of the systemic velocity of the inner binary) corresponds to 
the R\"{o}mer delay only, and the direct translation of the ETVs to the 
RVs can only be done when the two effects are separated, or the dynamical one 
is negligible.

We can clearly see the advantage of direct RV measurements of three stars, 
as the results obtained in this study are much more precise (1.5 per cent in $M_B$)
than from eclipse timing variations. However, the analysis of ETVs gives a slightly 
different set of orbital parameters, and takes into account more effects. Both 
approaches are therefore complementary, and together allow for a complete description of 
the system's dynamics and orbital architecture. With all masses directly and precisely 
measured, KIC~06525196 is a unique system, important to study formation and dynamical 
evolution of multiples.

\subsubsection{The tertiary of KIC~10191056}\label{sec_K1019_B}

\begin{figure}
\centering
\includegraphics[width=0.9\columnwidth]{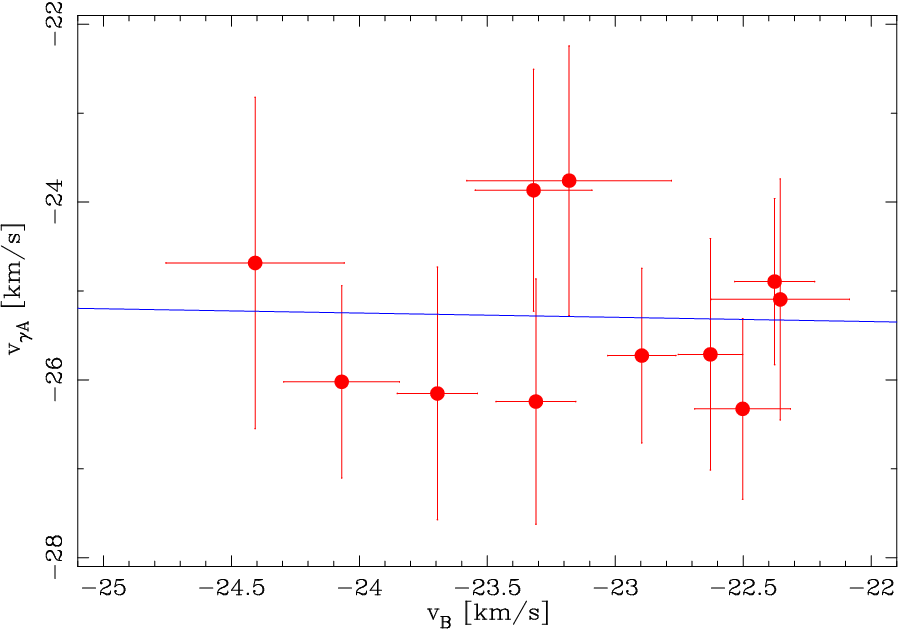}
\caption{A linear fit on the $v_B$/$v_\gamma$ plane, used to find
the mass ratio and systemic velocity of the KIC~10191056 triple.
}\label{fig_q_1019}
\end{figure}

\begin{figure}
\centering
\includegraphics[width=0.9\columnwidth]{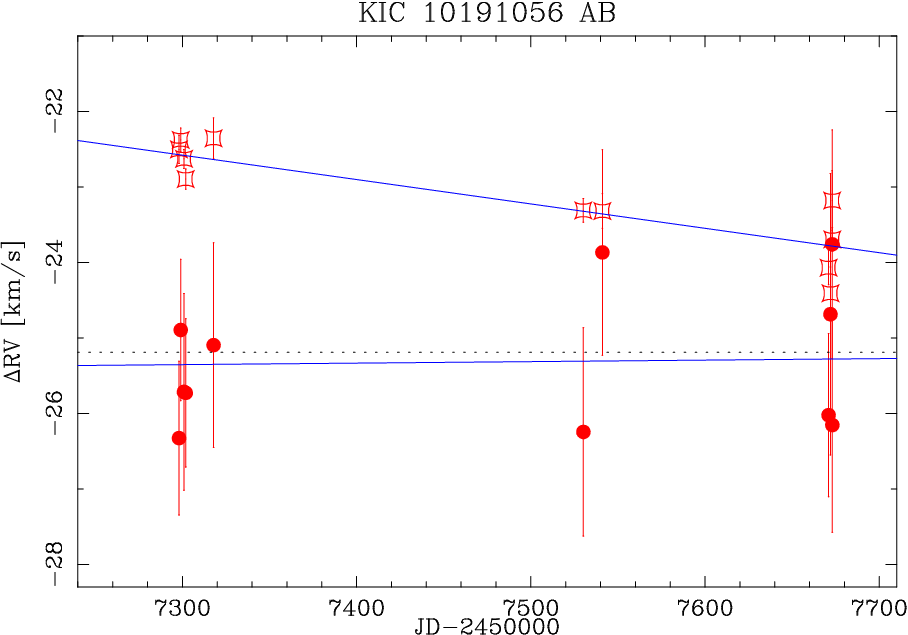}
\caption{A clear linear trend of the RVs of KIC~10191056~B, and a 
possible trend in systemic velocity of the eclipsing pair. 
Symbols are the same as in Fig.~\ref{fig_orb3_0652}.
The dotted line marks the probable systemic velocity of
the whole triple.
}\label{fig_orb3_1019}
\end{figure}

For the analysis of the inner eclipsing pair in this triple-lined system
we assumed that the two components resolved in
high-angular-resolution imaging are gravitationally bound and that the 
fainter star B is the source of the narrow spectral lines. We also assumed that
the change of its velocity ($v_B$) is linear, as is the change of the systemic
velocity of the inner pair A ($v_{\gamma,A}$). This is supported by the fact, that
in 33 years the position of B relatively to A has not changed much, and the
RVs of the tertiary are close to the systemic velocity of A.

We first calculated $v_{\gamma,A}$ for each observation, using the Equation \ref{eq_gam}.
Then we fitted a linear trend our measurements of $v_B$, obtaining
$\dot{v}_B=-0.00323(62)$~k\ms~d$^{-1}$ ($rms=0.35$~k\ms). We have then estimated
the mass ratio $q_{AB} = M_B/M_A = M_B/(M_1+M_2)$, and the systemic velocity of the 
whole system $\gamma$, by using a modified Equation~\ref{eq_gam}:
\begin{equation}
v_B = -q_{AB} v_{\gamma,A} + \gamma (1+q_{AB}),
\end{equation}
in which we substituted $v_2$ and $v_1$ with $v_B$ and $v_{\gamma,A}$, respectively.
The linear regression to this equation is shown in Figure~\ref{fig_q_1019}.
We found $\gamma=-25.2\pm0.9$~k\ms, and $q_{AB}=0.06\pm0.45$ ($rms=0.97$~k\ms), 
the mass of the tertiary $M_B$ is therefore $0.16\pm1.34$~M$_\odot$. This value is in
general agreement with the fact that the tertiary is fainter than each of the
eclipsing components, however seems to be too low for the observed magnitude differences,
and the formal uncertainty is very large due to the errors
of individual $v_{\gamma,A}$ measurements, coming from individual errors of 
$v_1$ and $v_2$. It is thus formally possible, that the tertiary is more massive, 
but other observational facts do not support this. We have, however, decided to keep
this conservative uncertainty in further analysis. Finally, we have estimated 
the scale of the linear trend $\dot{v}_{\gamma,A} = 0.0002\pm0.0014$~k\ms~d$^{-1}$.
The trend is, therefore, undistinguishable from zero. This is the value we used for 
the orbital fit described previously, and its uncertainty has been properly taken into account.
Our measurements of $v_{\gamma,A}$ and $v_B$, and both linear trends are shown
in Fig.~\ref{fig_orb3_1019}. However, because $\dot{v}_{\gamma,A}$ is indistinguishable
from zero, its confirmation requires further observations in the future.

\begin{figure}
\includegraphics[width=0.9\columnwidth]{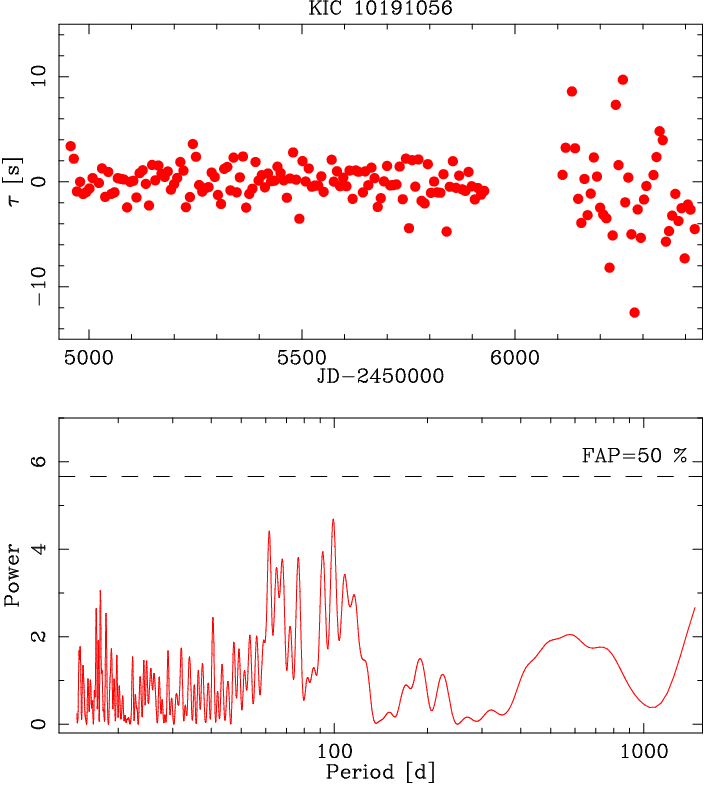}
\caption{{\it Top:} Our eclipse timing variations of KIC~10191056 as a function
of time. They do not show any detectable curvature 
{\it Bottom:} Their periodogram, which shows no significant peaks.
The horizontal line marks FAP = 50 per cent.
}\label{fig_etv_1019}
\end{figure}

Having $\dot{v}_{\gamma,A} = 0$ means that the tertiary is
in fact not gravitationally bound (a blend), or the period of the AB orbit
is too long to see the motion in the RVs. We mentioned before that the ETVs 
reported by \citet{gie15} show a small curvature, as well as a gradual diverging.
We have calculated our own ETVs, which we present in Figure~\ref{fig_etv_1019}.
Please note that the corresponding periodogram shows no statistically significant
peaks, meaning no short-period variations in KIC~10191056. We also do not see any 
curvature, down to the level of 2.5~s ($rms$ of our measurements)  
over the whole course of \kep observations, and even 1.4~s for data taken
before JD=2456000. The ETVs from \citet{gie15} vary by about 8 seconds, so we 
should be able to see the change. The reason why our results differ from those
of \citet{gie15} remains unclear to us.

Please also note that the measured visual separation between A and B has changed 
only by 0.16 arcsec in 33 years, but it can be explained by the absolute proper 
motion of the system, which is $7.1\pm1.6$~mas/yr \citep{gai16}.
If A and B are not gravitationally bound, or the period of their common orbit
is too long for a detection of any motion in RVs or ETVs, there might be another body 
causing the variation in the tertiary's velocity.
It might produce a linear trend, but there is also a possibility that the 
variation is short-period, and we have also investigated this scenario.

\begin{table}
\centering
\scriptsize
\caption{Parameters of four `planetary' orbital solutions to the RVs of
KIC~10191056~B.}\label{tab_v3_planet}
\begin{tabular}{lcccc}
\hline \hline
Parameter 		& Sol. 1	& Sol. 2	& Sol. 3	& Sol. 4 \\
\hline
$P$ (d)			&  9.892(25)	& 10.220(17)	& 17.887(37)	& 18.92(10)	\\
$T_P$ (JD-2457200)	& 97.1(1.2)	& 83.0(1.0)	& 97.66(44)	& 98.4(7.1)	\\
$K$ (k\ms)		&  1.05(20)	&  1.09(21)	&  0.95(22)	&  1.01(25)	\\
$\gamma$ (k\ms)	&-23.21(8)	&-23.08(7)	&-23.28(9)	&-23.50(17)	\\
$e$ 			&  0.39(26)	&  0.41(14)	&  0.65(21)	&  0.11(24)	\\
$\omega$ ($^\circ$)	& 247(22)	& 161(24)	& 272(17)	& 341(164) 	\\
$rms$ (\ms)		&    199	&    189	&    217	&    215	\\
$m\sin(i)$ (M$_J)^a$	& 10.2(2.3)	& 10.6(2.3)	&   9.3(2.8)	&  13.2(3.4)	\\
\hline
\end{tabular}
\\$^a$ Assuming mass of the host star =1~M$_\odot$.
\end{table}

\begin{figure}
\centering
\includegraphics[width=\columnwidth]{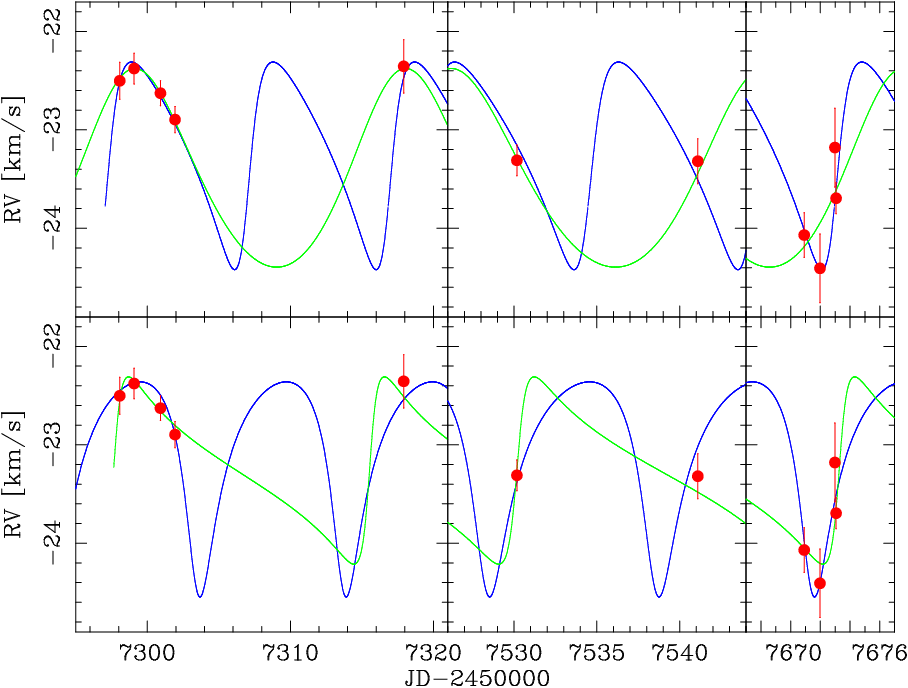}
\caption{Four `planetary' orbital fits to the RVs of the star B in 
KIC~10191056 (red points). Their parameters are listed in Tab.~\ref{tab_v3_planet}.
To improve clarity, in the upper row we show Sol. 1 (blue line) and 4 (green), 
and Sol. 2 (blue) and 3 (green) in the lower. Both rows have the same vertical scale.
}\label{fig_v3_planet}
\end{figure}

The two points with the highest values, and the `curvatures' seen in the first
and last four measurements, limit the possible periods to the range of $\sim$7--23 
days. We have explored this range of periods and found several values, that give 
satisfactory orbital fits. They are listed in Table~\ref{tab_v3_planet}, together 
with other (putative) orbital parameters, and also plotted on Figure~\ref{fig_v3_planet}. 
Please note that, except the last one, all solutions predict significant eccentricity. 
Assuming mass of the central star to be 1~M$_\odot$, all solutions lead to the lower mass 
limit of the companion in the planetary regime. Such a planet would be 
the first one found around a blend with (or a very wide companion to) an eclipsing 
binary. For the record, under the assumption of no linear trend,
the systemic velocity of the eclipsing pair is $\gamma_1=-25.2(5)$~k\ms,
and the two velocity amplitudes are exactly the same as in Table~\ref{tab_par_sb2}.

We would like to stress that these `planetary' solutions are highly uncertain 
due to small number of measurements. We also find the scenario with a linear 
trend in $v_B$ more probable, however our current data do not allow to
confirm it securely.

\section{Discussion}\label{sec_discus}

\subsection{Comparison with the TrES results}

\begin{figure}
\includegraphics[width=0.9\columnwidth]{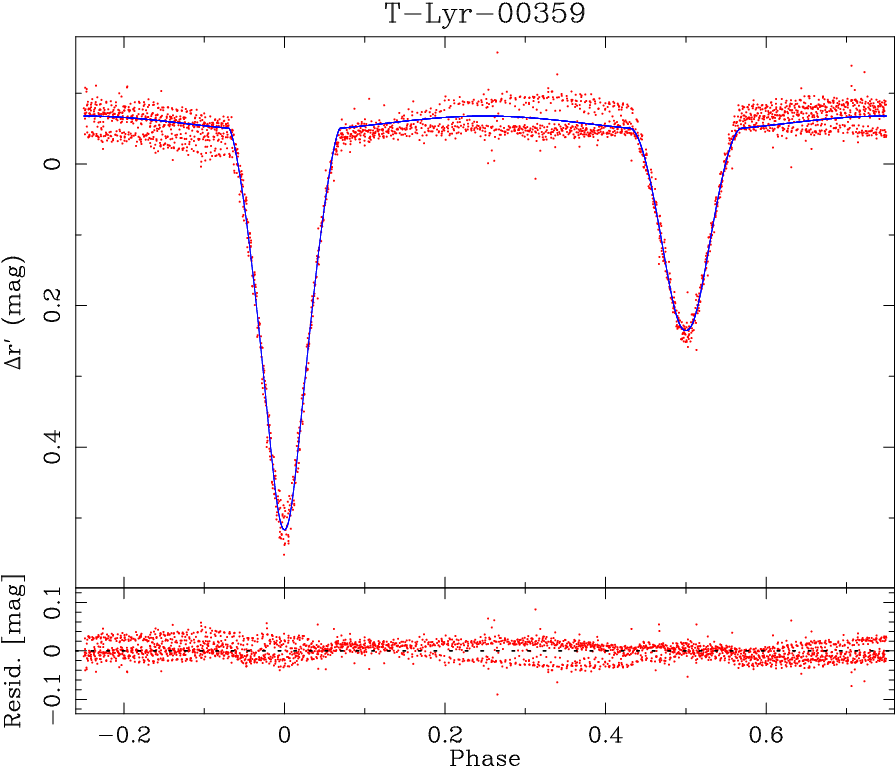}
\includegraphics[width=0.9\columnwidth]{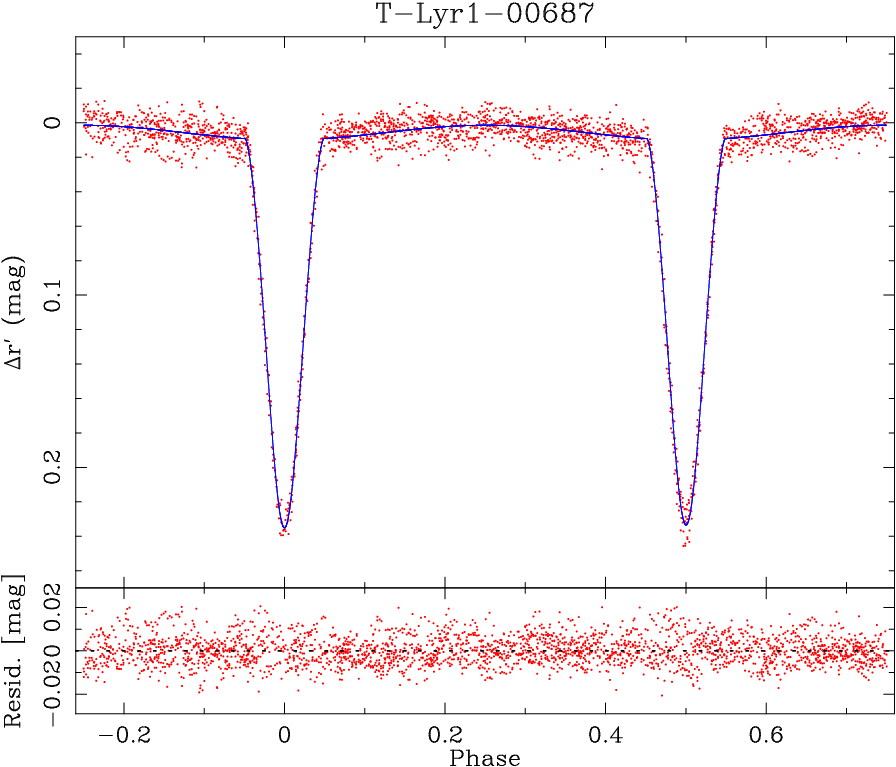}
\includegraphics[width=0.9\columnwidth]{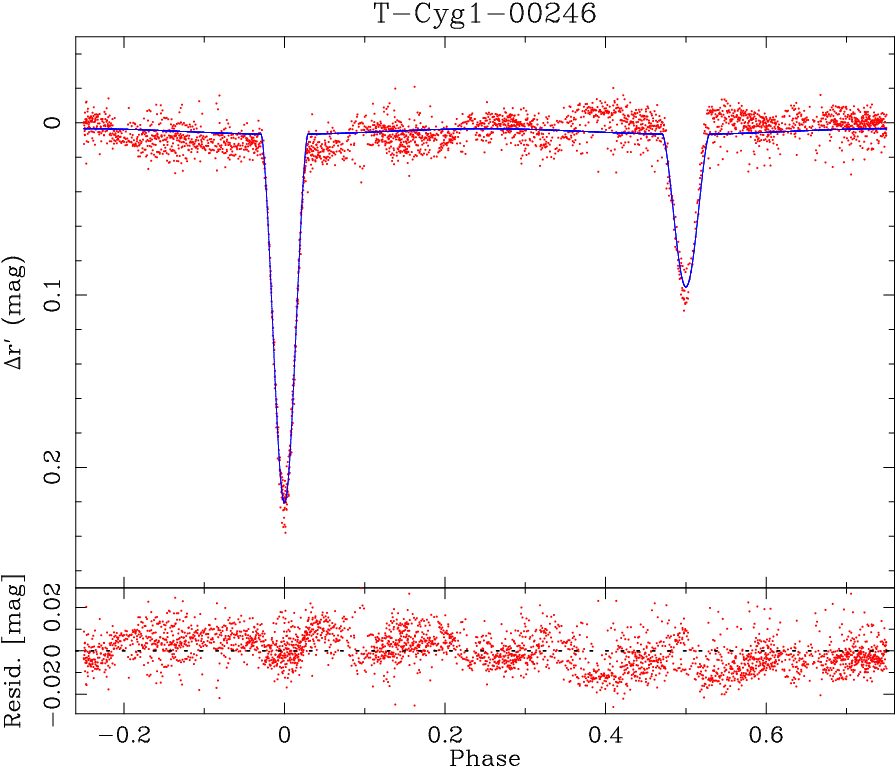}
\caption{Observed and modelled TrES light curves of 
T-Lyr1-00359 = KIC~08552540 (top),
T-Lyr1-00687 = KIC~10191056 (middle), and
T-Cyg1-00246 = KIC~11922782 (bottom).
Observations were made in the SDSS $r'$ band. 
Instrumental and/or spot-originated modulations are seen
outside of the eclipses in T-Lyr1-00359 and T-Cyg1-00246. 
}\label{fig_tres}
\end{figure}

\begin{table}
\centering
\caption{Results of the \jkt fit to the TrES light curves.
}\label{tab_tres}
\begin{tabular}{lccc}
\hline \hline
\multicolumn{1}{r}{KIC} & 08552540 & 10191056 & 11922782 \\
\multicolumn{1}{r}{T-}  & Lyr1-00359 & Lyr1-00687 & Cyg1-00246 \\
\hline
$P_{ecl}$ (d) & 1.061932(9) & 2.427610(34) & 3.512815(70)\\
$T_0$ (JD-2454900) & 54.102(12) & 55.0970(23) & 56.188(35) \\
$e$ & 0.0(fix) & 0.003(fix) & 0.0(fix) \\
$\omega$ ($^\circ$) & --- & 278(7) & --- \\
$r_1$ & 0.2426(42) & 0.175(13) & 0.127(21) \\
$r_2$ & 0.1841(54) & 0.154(15) & 0.084(22) \\
$i$ ($^\circ$) & 85.65(98) & 82.24(37) & 83.9(1.2) \\
$J$ & 0.91(11) & 0.968(28) & 0.401(25) \\
$L_2/L_1$ & 0.305(13) & 0.77(30) & 0.19$^{+0.45}_{-0.08}$ \\
$L_3/L_{tot}$ & 0.0(fix) & 0.0(fix) & 0.16$^{+0.23}_{-0.16}$\\
$rms_{LC}$ (mmag) & 18.0 & 6.6 & 7.9\\
\hline
\end{tabular}
\end{table}

\begin{table}
\centering
\caption{Comparison of results obtained in this study 
(from HIDES and \kep data) for three TrES systems with results from 
\citet{dev08}.
}\label{tab_dev}
\begin{tabular}{lcc}
\hline \hline
Parameter	& This work & \citet{dev08} \\
\hline
\multicolumn{3}{c}{\it T-Lyr1-00359 = KIC~08552540}\\
$P_{ecl}$ (d)		& 1.06193441(4)	& 1.061922(15) \\
$M_1$ (M$_\odot$)	& 1.153(36)	& 1.655(15) \\
$M_2$ (M$_\odot$)	& 0.956(28)	& 1.296(12) \\
\multicolumn{3}{c}{\it T-Lyr1-00687 = KIC~10191056}\\
$P_{ecl}$ (d)		& 2.427494881(19) & 2.427512(79) \\
$M_1$ (M$_\odot$)	& 1.575(30)	& 1.209(13) \\
$M_2$ (M$_\odot$)	& 1.420(36)	& 1.208(13) \\
\multicolumn{3}{c}{\it T-Cyg1-00246 = KIC~11922782}\\
$P_{ecl}$ (d)		& 3.5129340(4)	& 3.51306(16) \\
$M_1$ (M$_\odot$)	& 1.065(10)	& 1.498(26) \\
$M_2$ (M$_\odot$)	& 0.835(6)	& 0.970(32) \\
\hline
\end{tabular}
\end{table}

Three of our systems have additional photometric data available from
the TrES survey, and the catalogue of eclipsing binaries prepared
by \citet{dev08}. We have fitted model LCs to these data, using the
\jkt code in almost the same way as for the \kep photometry 
(the differences were that the TrES LCs were not divided into smaller
subsets for the RS stage, the small eccentricity of KIC~10191056
was held fixed, and third light of KIC~11922782 was fitted for,
due to much shallower eclipses in the TrES curve). 
The results are listed in Table~\ref{tab_tres}.
For most of the parameters independent from the bandpass (i.e. 
$P,e,\omega,i,r_1,r_2$) the agreement is well within the given 1$\sigma$
uncertainties. The results from \kep photometry are also significantly
more precise, with the exception of KIC~08552540, which is the system
with the largest spot-originated brightness variations in the sample,
and for which the \kep-based results are only slightly better. This case
shows that even with such an exquisite instrumental precision, the 
resulting parameters may still not be derived as well as expected, unless
the additional variability is removed carefully and correctly. 

Notably, in case of KIC~11922782 the $rms$ is essentially the same for TrES and 
\kep LCs, but the TrES results were hampered by the third light, and the fact 
that over the course of the observations the spot-originated modulation did not 
change much, while it did in \kep observations, and therefore could be averaged
over the orbital phase. The TrES LC clearly shows systematic residuals 
(Fig.~\ref{fig_tres}, bottom). An alternative explanation for shallower TrES
eclipses may be a change in orbital inclination. We, however, find it unlikely, 
despite such cases being observed by \kep \citep{sla11,rap13}. The TrES 
measurements were done with a relatively large 30''$\times$30'' photometric 
aperture, and a different, i.e. smaller inclination would produce eclipses that 
last shorter, but we don't see such effect. The best fit was found for third light 
contribution of over 15 per cent, although with a very large uncertainty.

In their work \citet{dev08} also used their LCs to derive stellar masses.
They used the \textsc{meci} code \citep{dev06}, which looks for the most
probable combination of two masses and age of the system, by comparing
global photometric properties (brightness and colours) with a set of
isochrones. Having our HIDES spectroscopy and RVs, we can compare our direct
determinations of masses with their indirect results. Such comparison is
shown in Table~\ref{tab_dev}. One can quickly note that the masses found 
indirectly are in a strong disagreement with our findings. This is, 
however, not the first example of such a discrepancy that can be found in
literature. The same \textsc{meci} method (with small modifications) was used later 
for example by \citet{hel13} on a sample of DEBs from the Galactic bulge, 
or by \citet{lee15} on ASAS, NSVS, and LINEAR systems. While the Galactic 
bulge systems do not have their spectroscopy done so far, many of the ASAS 
ones from \citet{lee15} have their parameters published 
\citep[e.g.][]{dim14,dju11,hel09,hel15a,rat13,rat16,roz13}.

There are also many systems whose light curves alone were 
analysed with other codes, like \textsc{phoebe} \citep{prs05}. This code
works on Kopal modified potentials, which are dependent on the mass ratio, 
and determine the shape of the out-of-eclipse ellipsoidal variations. 
It is therefore theoretically possible to estimate $q$ for systems showing 
this phenomenon. In few cases, a complete LC+RV analysis have also been
performed, and masses and/or their ratios found both ways can also be compared,
like KIC~06525196 in this work vs. \citet{rap13}.

Such a comparison between the direct and indirect measurements is not the main scope 
of this paper. Here we only conclude, that 
there is very little agreement between direct and indirect mass and mass ratio
determinations, and if there is, it is either accidental, or due to very large
uncertainties of the indirect one. This means that the LC-only results are 
very insecure and should be treated with a lot of caution.

\subsection{Age, evolutionary status, $T_{eff}$, and distance \label{sec_age}}

\begin{figure*}
\includegraphics[width=0.32\textwidth]{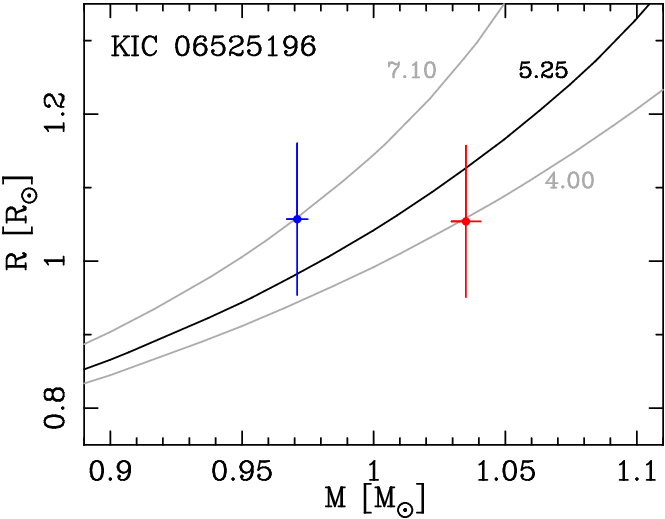}
\includegraphics[width=0.32\textwidth]{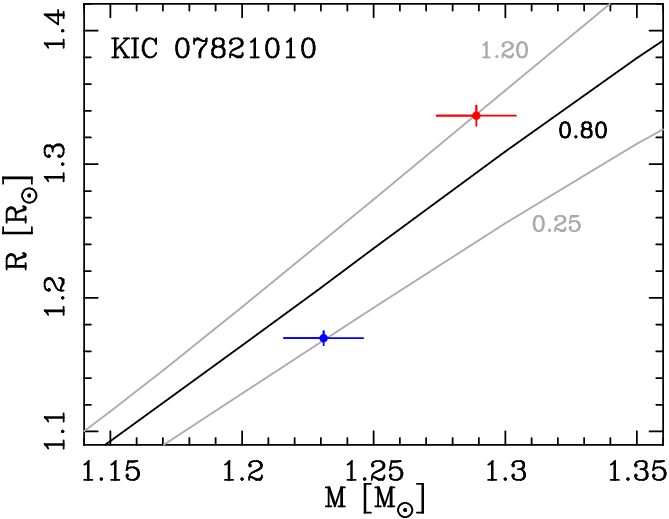}
\includegraphics[width=0.32\textwidth]{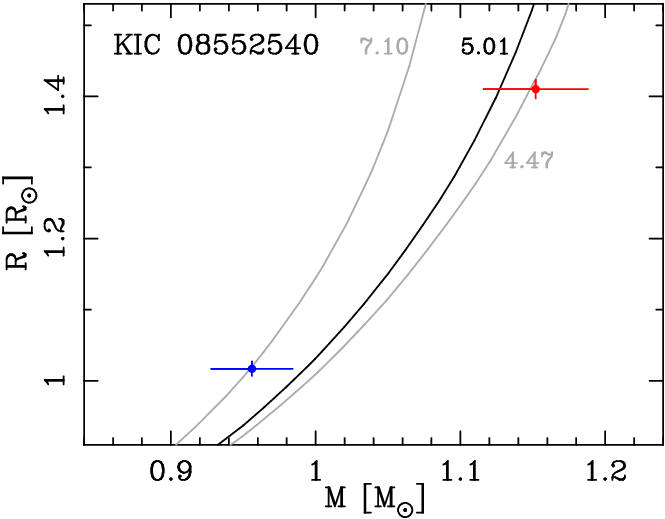}
\includegraphics[width=0.32\textwidth]{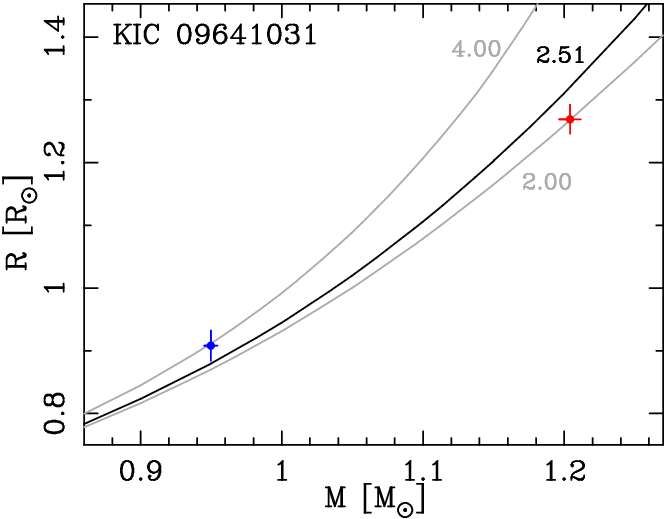}
\includegraphics[width=0.32\textwidth]{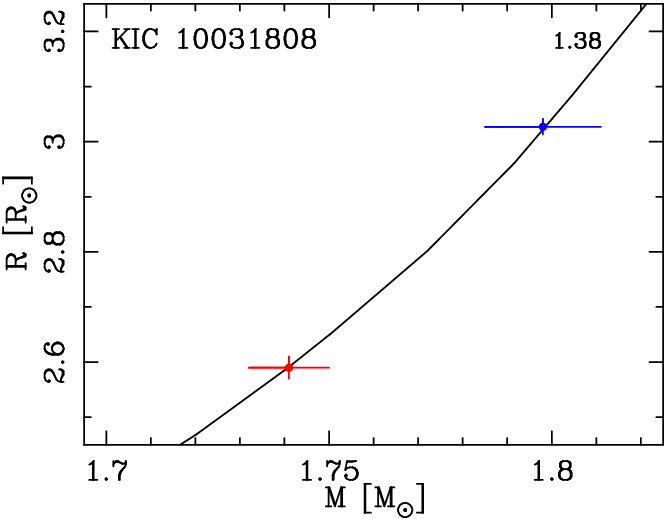}
\includegraphics[width=0.32\textwidth]{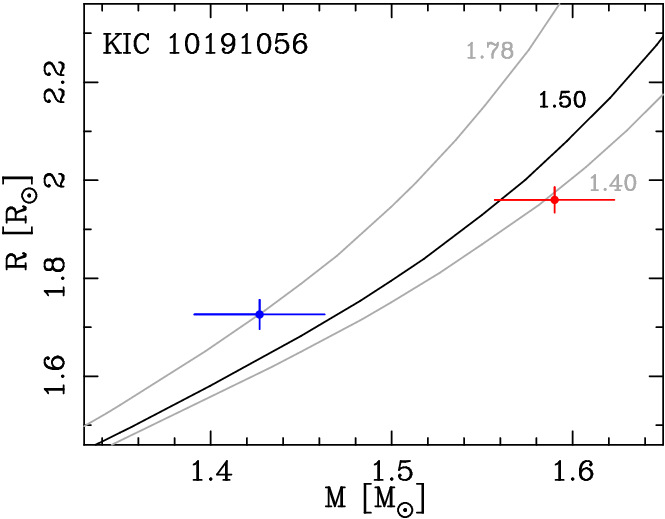}
\includegraphics[width=0.32\textwidth]{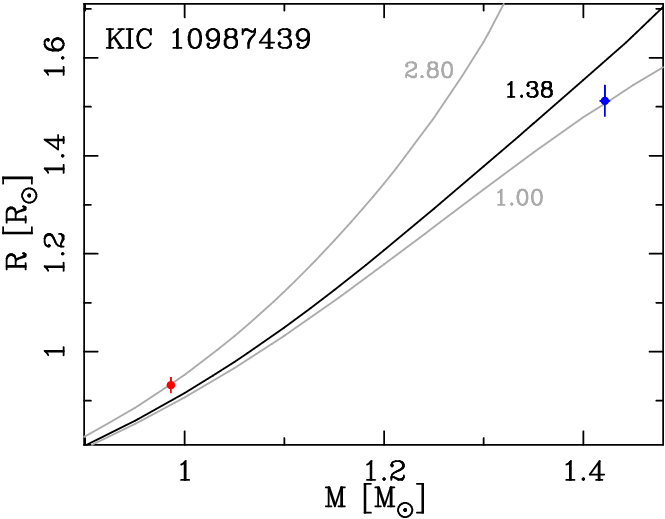}
\includegraphics[width=0.32\textwidth]{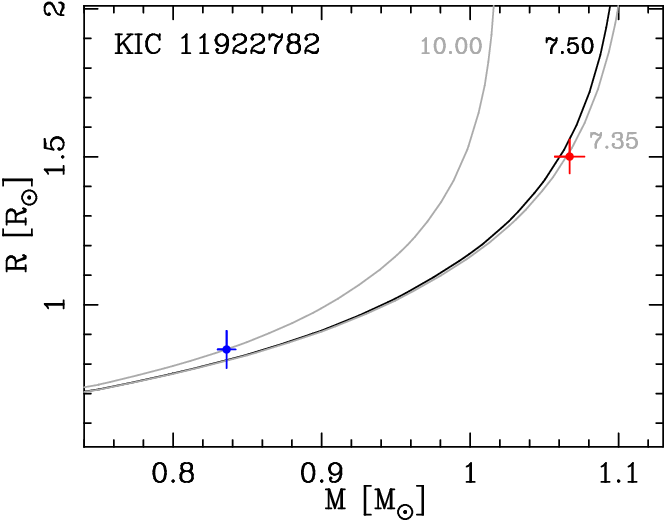}
\caption{Comparison of our results with PARSEC solar metallicity
isochrones on the mass-radius plane. Primaries are shown with 
red, and secondaries with blue symbols, and with 1$\sigma$ errorbars.
Black lines are isochrones that best match both components simultaneously,
while grey lines match one of the components (if different from the 
simultaneous best match). Ages in Gyr are also given.
}\label{fig_iso}
\end{figure*}

In this Section we asses the age and evolutionary status
of each system by comparing our results, namely masses and radii,
with the theoretical PARSEC isochrones \citep{bre12}, that include
calculation of absolute magnitudes in the \kep photometric band. From the 
best-matching isochrone we estimate the effective temperatures, and use them to
infer distances with \textsc{jktabsdim}. We compare these distances with the
recently published {\it Gaia} Data Release 1 \citep[GDR1;][]{gai16}, which 
gives results of the {\it Tycho-Gaia} Astrometric Solution \citep{mic15}.

In Figure~\ref{fig_iso} we present our results of isochrone fitting on mass-radius 
planes. We'd like to remind that these result assume solar metallicity, 
therefore the ages should be treated with some caution, because of the age-metallicity
degeneration. These results could be corrected if precise multi-band light
curves were obtained, and system's metallicity and temperatures of both components 
independently estimated.

\subsubsection{KIC~06525196}
It is a problematic system for two reasons. First, it is a triple and we only 
have information of fractional fluxes in the \kep band, not in $B,V,J,H,K$ 
(no multi-band photometry). Therefore we could not calculate the distance
with the \jktabs code. Fortunately, we know the mass of the third star precisely, 
(with 1.5 per cent relative precision, see Tab.~\ref{tab_par_3rd})
which helped to put some additional constraints.
Second, the relatively large uncertainties of radii, and the fact that in our solution
the less massive secondary turned out to be formally larger, caused that
two scenarios were possible for this system: `young' with the age of 20~Myr 
($\log(\tau)=7.30$) indicating the pre-main-sequence stage, and `old' with the age of
5.25~Gyr ($\log(\tau)=9.72$). Comparison of predicted flux ratios for all three stars 
(from absolute magnitude differences) favoured the `old' one, however the agreement 
was only marginal. In this scenario, the primary and secondary lay on isochrones
of 4.0~Gyr ($\log(\tau)=9.60$) and 7.1~Gyr ($\log(\tau)=9.85$), respectively.

The predicted temperatures are $6000\pm100$, $5800\pm100$, and $4850\pm100$~K
for the primary, secondary, and tertiary, respectively. The respective absolute
magnitudes in the \kep filter ($Kmag$) are: 4.30, 4.75, and 6.23~mag, with 
0.20~mag uncertainty for all components. Large errors in radii lead to large 
uncertainties in absolute magnitudes. Also the error of $L_2/L_1$ from our \jkt 
solution is significant. We suspect that its value, as well as the value of 
$L_3$, might be overestimated. The predicted total $Kmag$ is 
$3.65\pm0.22$~mag. By comparing it with the observed $kmag=10.154$, we obtain
the distance $d=200\pm20$~pc, assuming no extinction. Taking this value, and 
the physical size of the major semi-axis $a_{AB}=1.539\pm0.010$~AU, we can
calculate its angular size: $\hat{a}_{AB}=7.7\pm0.8$~mas. The resulting 
distance is in a very good agreement with the GDR1 value of $205\pm14$~pc,
which confirms the correctness of our solution and usage of the solar-metallicity
isochrone. Notably, with its physical parameters, and probable metallicity and age, 
the primary can be considered a solar analogue. 

\subsubsection{KIC~07821010} 
Comparison of our results with solar metallicity 
PARSEC isochrones on the $M/R$ plane shows that both components are on the main 
sequence. The isochrone that fits best to the whole system is found for the age of 
$\sim$800~Myr ($\log(\tau)=8.9$). 
The primary and secondary components 
lay on isochrones of 1.20~Gyr ($\log(\tau)=9.075$) and 250~Myr ($\log(\tau)=8.40$), 
respectively. The predicted effective temperatures are $T_{eff1}=6600\pm100$ and 
$T_{eff2}=6450\pm200$~K. We use them and the available $BVJHK$ photometry as 
the input for {\sc jktabsdim} to estimate the $E(B-V)$ and distance. 
We found that the best consistency between various bands is found for 
$E(B-V)\simeq 0.06$, and the resulting distance is $347\pm14$~pc. The GDR1
distance -- $446\pm183$~pc -- is in a formal agreement, however significantly
less precise.

\subsubsection{KIC~08552540 (V2277 Cyg)}
This system turns out to be relatively old, and is a good example
of how fast rotation (due to tidal locking) can help sustain a high level of
activity in mature, solar-like stars.
The comparison with isochrones suggests the age of 5.01~Gyr ($\log(\tau)=9.70$).
The primary and secondary lay on isochrones of 4.47~Gyr ($\log(\tau)=9.65$)
and 7.1~Gyr ($\log(\tau)=9.85$), respectively. Their predicted temperatures
are $6200\pm100$ and $5700\pm200$~K. The distance, as calculated by the
\jktabs code, is $d=262\pm7$~pc, with no extinction. The small error
comes from the very consistent values individually calculated for each band.

The \jkt analysis of $V$ and $I$-band curves from ASAS-K, gives respective
observed magnitudes of $10.668\pm0.024$ and $9.972\pm0.013$~mag for the 
primary, and $11.955\pm0.085$ and $11.230\pm0.055$~mag for the secondary. 
This leads to the $(V-I)$ colours of $0.70\pm0.03$ and $0.73\pm0.10$~mag 
for the primary and secondary, respectively.
The second value has its uncertainty too large to be useful to constrain the
temperature. The first one alone may be used to assess the interstellar
reddening. The 7.1~Gyr PARSEC isochrone gives $(V-I)=0.669$~mag for the
primary's mass, which is close to, but not exactly the value we got from the ASAS-K data. 
The $E(V-I)$ is therefore 0.031~mag, which translates into $E(B-V)=0.043$~mag. 
By putting this value in \jktabs we obtain the distance of $d=257\pm10$~pc, 
which is in a very good agreement with our previous value, but has larger uncertainty.

However, the GDR1 distance is $227\pm17$~pc, so the agreement is only in a 
2$\sigma$ level. If a different metallicity isochrone was used, the predicted
temperatures could have been lower, resulting in a smaller \jktabs distance.

\subsubsection{KIC~09641031 (FL Lyr)}
Our mass and radius measurements are best reproduced by a 2.51~Gyr
($\log(\tau)=9.40$) isochrone. The agreement is at a $\sim$1$\sigma$
level, and is better than if the original results of \citet{pop86}
were compared with the PARSEC set. Individually, the components lay
on isochrones of 2.00 ($\log(\tau)=9.30$, primary) and 4.00~Gyr
($\log(\tau)=9.60$, secondary). 

This is a special case in our sample, as it is the only system with 
independent estimates of the effective temperatures: $6150\pm100$ and 
$5300\pm100$~K for the primary and secondary, respectively \citep{pop86}.
We used them together with total system's brightness in $B,V,J,H,K$ as an 
input in {\sc jktabsdim}, and calculated the distance. We obtained the 
weighted average value of $d=131\pm4$~pc (no extinction), which is in excellent 
agreement with the {\it Hipparcos} value of $130.2\pm12.5$~pc \citep{vLe07},
and only marginal with the GDR1 value of $138\pm4$~pc.
Parameters from the original solution of \citet{pop86} give 
a slightly higher value of $d=134\pm4$~pc. This may be an indication 
that our solution is less accurate.

For the record, the temperatures predicted by the best-fitting isochrone 
are $6500\pm150$ and $5600\pm100$~K for the primary and secondary, 
respectively. A good consistency in distance calculation is reached for
$E(B-V)\simeq0.09$, which results in $d=133\pm5$~pc -- closer to the GDR1
value than for the Popper's temperatures. In any case, 
despite a significantly different temperature scale, all the results
are still in agreement with the ones given above. The consistency
between distances calculated for different passbands is also very high. 
To confirm which distance is the correct one, the {\it Gaia}
satellite will have to provide its determination at the level of 
$\sim$1~pc, or $<0.1$~mas in parallax, which can be expected from the 
future data releases.

\subsubsection{KIC~10031808} 
Comparison of our results with the solar metallicity PARSEC isochrone on
the $M/R$ plane shows a very good agreement for an age of 1.38~Gyr 
($\log(\tau)=9.14$) -- both components are located on this line. 
Both are already somewhat evolved and are about to leave the main sequence,
which makes this system important for studying late phases of main sequence
evolution. In this stage the age-metallicity degeneracy is weaker than for 
earlier ages, as the slope of the isochrone on the $M/R$ plane changes
with the metal content. With masses and radii precise enough, one could
constrain [$M/H$] quite securely. In this case, however, the agreement is 
already excellent, meaning that the system's true metallicity is close to solar.
The model predicts temperatures of $6850\pm100$ and $7100\pm100$~K for 
the primary and secondary, respectively. When incorporated into 
{\sc jktabsdim}, they result in $d=442\pm15$~pc for $E(B-V)\simeq0.15$~mag. 
The GDR1 value of $500\pm73$~pc agrees quite well, however mainly because of 
the large uncertainty.

One should note that the temperatures, and values of $\log(g)$ 
(Tab.~\ref{tab_par_sb2}) place both components in $\delta$~Scuti and/or 
$\gamma$~Doradus instability strips \citep{kah16}. As shown in 
Section~\ref{res_1003}, we identified many modes of $\gamma$~Dor type 
pulsations in the residuals of the LC fit, while $\delta$~Scuti are not
seen. We can therefore conclude that at least one component is a $\gamma$~Dor 
variable. This makes KIC~10031808 even more interesting, as examples of such 
stars in eclipsing systems, especially with well measured parameters, are very 
rare \citep[e.g.][]{deb13,guo16}.

\subsubsection{KIC~10191056}
The two eclipsing components of this triple
seem to be coming to an end of their main sequence evolution. The system's 
parameters are consistent with a 1.50~Gyr ($\log(\tau)=9.175$) isochrone, 
while the the primary and secondary are better individually represented
by 1.40~Gyr ($\log(\tau)=9.15$) and 1.78~Gyr ($\log(\tau)=9.25$)
lines, respectively.

The predicted temperatures are $T_{eff,1}=7100\pm150$~K, and 
$T_{eff,2}=6900\pm100$~K. We can use them in \jktabs to estimate the distance
to the system, but only with magnitudes in $V$ and $I$, as the ASAS-K data allow 
us to estimate the tertiary's contribution only in those bands.
We found that the $V$ and $I$ observed magnitudes of the three components are:
$11.84\pm0.26$ and $11.07\pm0.08$~mag for the primary,
$12.07\pm0.33$ and $11.49\pm0.12$~mag for the secondary, and
$12.55\pm0.43$ and $11.85\pm0.17$~mag for the third star.
Unfortunately, the resulting $(V-I)$ colours -- $0.77\pm0.27$, $0.58\pm0.35$, and 
$0.70\pm0.39$~mag for the primary, secondary and tertiary, respectively -- are not 
precise enough to put any constraints on the effective temperatures.

After correcting for the third light, and assuming no interstellar extinction,
we obtain the distance of $d=688\pm38$~pc, which is the largest one in our sample.
When the distance is estimated from the observed contributions to the flux in
the \kep band, and compared with absolute values from the isochrone, we obtain
a similar value of $700\pm45$~pc. If we want to force the distances
calculated in \jktabs for $V$ and $I$-band individually to be equal, we must 
assume $E(B-V)=0.23$~mag. In such case, we obtain $d=550\pm60$~pc. Comparison
with the GDR1, which gives $562\pm112$~pc, suggests that this is a better approach,
but the large uncertainty makes it not completely conclusive. However, no 
interstellar extinction at such a distance seems unlikely.

\subsubsection{KIC~10987439}
This system is interesting as it has the smallest mass ratio (when defined as
the smaller mass over the larger) in our sample. In principle it is more 
difficult to fit a single isochrone to two points laying far from each other
on the $M$--$R$ or Hertzsprung-Russell (H-R) diagram. In this case there is no
solar-metallicity isochrone that matches both components at a 1$\sigma$ level.

These stars seem to reside on the main sequence. The best match to both of them 
simultaneously was found for the age of 1.38~Gyr ($\log(\tau)=9.10$).
The primary and secondary separately are best reproduced by the ages of
2.8~Gyr ($\log(\tau)=9.45$) and 1.0~Gyr ($\log(\tau)=9.00$), respectively.
The system is not active, therefore this large discrepancy (note small mass
and radius errors) can not be attributed to spots. We presume that the two 
components have different values of the mixing length parameter $\alpha$. 
Such a solution has been proposed to explain the observed properties of several 
F,G,K-type stars in eclipsing binaries \citep[e.g.][]{cla09,vos12}.

The best-fitting isochrone predicts effective temperatures of $5700\pm200$ and
$6950\pm100$~K. This results in the \jktabs distance of $334\pm12$~pc, at
$E(B-V)=0.1$~mag. In this case the agreement with GDR1, which gives $348\pm24$~pc,
is also very good.

\subsubsection{KIC~11922782}
The last system in our sample is also the oldest one. The observed masses and
radii are very well reproduced by a 7.50~Gyr isochrone ($\log(\tau)=9.875$),
while primary and secondary formally lay on isochrones of 7.35~Gyr ($\log(\tau)=9.87$)
and 10.0~Gyr ($\log(\tau)=10.0$), respectively. The former has already evolved out of 
the main sequence, and constrains the age much stronger than the secondary. With the
mass very close to 1~M$_\odot$, but a significantly larger radius, it can serve
as an insight into the state of our Sun in few billion years. Despite relatively
large uncertainty in radius, the secondary seems to follow the trend observed in
many low-mass, active components of short-period eclipsing binaries, that the
observed radius is larger than expected from isochrones.

The predicted temperatures are $5900\pm100$ and $5250\pm100$~K for the primary
and secondary, respectively. The distance is therefore $225\pm12$~pc, at
$E(B-V)=0.07$~mag. There is no solution for this system in the GDR1.

\section{Summary}
We have studied six double- and two triple-lined detached eclipsing binaries
from the KEBC database, for which we obtained absolute values of masses and radii,
and estimated other parameters from the comparison with theoretical isochrones.
We also conclude that stellar masses determined indirectly (without RVs), with the 
exception of asteroseismology, are not reliable. Eclipse timing variation analysis of 
selected targets was also performed. 

In some cases we have managed to reach precision good enough to test the models.
The PARSEC models managed to reproduce our results quite well. Each 
best-fitting isochrone agreed with our masses and radii within 3$\sigma$. 
The differences can be attributed to the true metallicities being different
from solar, and/or to different mixing length parameters for each component 
of a given pair. Also the estimated distances were in good agreement with those 
from GDR1, which for the largest distances turned out to be much less precise 
than our results. 

In this work we managed to identify interesting targets, worth further studies.
In our sample we see various classes of objects, like low-mass stars (e.g. KIC~11922782),
a $\gamma$~Dor pulsator (KIC~10031808), or a hierarchical triple with nearly
complete dynamical description (KIC~0652196~AB). Further work on these objects
should focus on spectroscopic analysis, in order to independently determine
temperatures and metallicities, multi-colour photometry, and precise correction
of brightness variations (spots, pulsations) in order to improve the 
precision in radii. Also, a long-term spectroscopic monitoring of the unsolved
case of KIC~10191056~B, and a detailed frequency analysis of KIC~10031808 are
encouraged.

\section*{Acknowledgments}
We would like to thank the anonymous Referee and Prof. Andrzej Pigulski from 
the Astronomical Institute of the Wroc{\l}aw University, for fruitful discussions 
and valuable suggestions.

This research has made use of the SIMBAD database, operated at CDS, Strasbourg, France.
This work has made use of data from the European Space Agency (ESA)
mission {\it Gaia} (\url{http://www.cosmos.esa.int/gaia}), processed by
the {\it Gaia} Data Processing and Analysis Consortium (DPAC,
\url{http://www.cosmos.esa.int/web/gaia/dpac/consortium}). Funding
for the DPAC has been provided by national institutions, in particular
the institutions participating in the {\it Gaia} Multilateral Agreement.

KGH acknowledges support provided by the National Astronomical Observatory 
of Japan as Subaru Astronomical Research Fellow.
This work is supported by the Polish National Science Center grants 
2011/03/N/ST9/03192 and 2015/16/S/ST9/00461, by
the European Research Council through a Starting Grant, by the
Foundation for Polish Science through `Idee dla Polski' funding
scheme, and by the Polish Ministry of Science and Higher Education
through grant W103/ERC/2011.

\appendix

\section{RV measurements}
In Table \ref{tab_RV} we present individual RV measurements used in this work.
Both KIC~06525196 and 10191056 have been treated as two SB2s, and the systemic
velocities of their inner pairs are given in Tab. \ref{tab_RV} as $v_1$. 
Time stamp is BJD-2450000. Exposure times and $S/N$ around $\lambda=5500$~\AA\,are
also given.

\begin{table*}
\centering
\caption{HIDES radial velocities. Complete Table is available
 in the on-line version of the manuscript.}\label{tab_RV}
\begin{tabular}{lcccclcc}
\hline \hline
BJD & $v_1$ & $\epsilon_1$ & $v_2$ & $\epsilon_2$ & KIC & $t_{exp}$ & $S/N$ \\
-2450000 & (k\ms ) & (k\ms ) & (k\ms ) & (k\ms )  & & (s) & \\
\hline
6865.062301 &   58.509 & 0.232 &  -35.422 & 0.312 & 6525196 A  &  900 & 54 \\
6865.220925 &   77.453 & 0.347 &  -56.066 & 0.399 & 6525196 A  & 1200 & 66 \\
6866.000781 &   77.885 & 0.362 &  -56.706 & 0.436 & 6525196 A  & 1200 & 43 \\
... \\
6865.062301 &   13.018 & 0.218 &   ---    &  ---  & 6525196 AB &  900 & 54 \\
6865.220925 &   12.789 & 0.301 &  -16.542 & 0.052 & 6525196 AB & 1200 & 66 \\ 
6866.000781 &   12.702 & 0.317 &  -16.415 & 0.070 & 6525196 AB & 1200 & 43 \\ 
... \\
6867.030540 &  -46.473 & 0.092 &   13.313 & 0.145 & 7821010    & 1800 & 48 \\
6869.142604 &  -38.627 & 0.361 &    5.871 & 0.532 & 7821010    & 1500 & 23 \\
6914.079695 &  -51.604 & 0.102 &   18.896 & 0.132 & 7821010    & 1500 & 63 \\
... \\
\hline
\end{tabular}
\end{table*}

\section{Eclipse timing variation measurements}
In Table \ref{tab_etv} we show our own measurements of
eclipse timing variations for three system, as derived with the 
method described in Section \ref{sec_etv}.

\begin{table}
\centering
\caption{The eclipse timing variations used in this work ($\tau$)
and their uncertainties $\epsilon$, calculated with the method of
\citet{koz11}. Only a portion of the table 
is shown here, and the complete one is available on-line.}\label{tab_etv}
\begin{tabular}{lccl}
\hline \hline
BJD & $\tau$ & $\epsilon$ & KIC \\
-2450000 & (s) & (s) &\\
\hline
4958.4024415 &  -44.3 &  15.4 & 06525196 \\
4968.8342515 &  -71.7 &  20.2 & 06525196 \\
4978.0706240 & -120.4 &  19.3 & 06525196 \\
... \\
4957.2275491 &  -3.2 & 12.7 & 09641031 \\  
4964.5839094 &   2.1 & 14.8 & 09641031 \\   
4971.9402599 &   5.0 & 10.2 & 09641031 \\   
... \\
4957.2277765 &   3.6 &  8.4 & 10191056 \\
4964.5841102 &   2.4 &  7.5 & 10191056 \\
4971.9404315 &  -0.7 &  6.8 & 10191056 \\
... \\
\hline
\end{tabular}
\end{table}

\label{lastpage}

\end{document}